\documentclass{jfm}

\usepackage{graphicx}
\usepackage{natbib}
\usepackage{color}
\ifCUPmtlplainloaded \else
  \checkfont{eurm10}
  \iffontfound
    \IfFileExists{upmath.sty}
      {\typeout{^^JFound AMS Euler Roman fonts on the system,
                   using the 'upmath' package.^^J}%
       \usepackage{upmath}}
      {\typeout{^^JFound AMS Euler Roman fonts on the system, but you
                   dont seem to have the}%
       \typeout{'upmath' package installed. JFM.cls can take advantage
                 of these fonts,^^Jif you use 'upmath' package.^^J}%
      }
  \else
  \fi
\fi

\ifCUPmtlplainloaded \else
  \checkfont{msam10}
  \iffontfound
    \IfFileExists{amssymb.sty}
      {\typeout{^^JFound AMS Symbol fonts on the system, using the
                'amssymb' package.^^J}%
       \usepackage{amssymb}%
         \let\leq=\leqslant
         
      }{}
  \fi
\fi

\ifCUPmtlplainloaded \else
  \IfFileExists{amsbsy.sty}
    {\typeout{^^JFound the 'amsbsy' package on the system, using it.^^J}%
     \usepackage{amsbsy}}
    {\providecommand\boldsymbol[1]{\mbox{\boldmath $##1$}}}
\fi

\providecommand\bnabla{\boldsymbol{\nabla}}
\providecommand\bcdot{\boldsymbol{\cdot}}

\newsavebox{\astrutbox}
\sbox{\astrutbox}{\rule[-5pt]{0pt}{20pt}}

\title[Lagrangian and Eulerian analysis]{Statistical properties of supersonic turbulence in the Lagrangian and Eulerian frameworks}
\author[L. Konstandin, C. Federrath, R. Klessen and W. Schmidt]
{L\ls U\ls K\ls A\ls S\ns K\ls O\ls N\ls S\ls T\ls A\ls N\ls D\ls I\ls N$^1$\ls,
 C\ls H\ls R\ls I\ls S\ls T\ls O\ls P\ls H\ns F\ls E\ls D\ls E\ls R\ls R\ls A\ls T\ls H$^{1,\,2,\,3}$\ls,
 R\ls A\ls L\ls F\ns S.\ns K\ls L\ls E\ls S\ls S\ls E\ls N$^{1}$\ns
 \and W\ls O\ls L\ls F\ls R\ls A\ls M\ns S\ls C\ls H\ls M\ls I\ls D\ls T$^4$}
\affiliation{$^1$ Zentrum f\"{u}r Astronomie der Universit\"at Heidelberg, Institut f\"{u}r Theoretische Astrophysik, Albert-Ueberle-Str. 2, D-69120 Heidelberg, Germany\\[\affilskip]
$^2$ Ecole Normale Sup\'{e}rieure de Lyon, Centre de Recherche Astrophysique, 46 All\'{e}e d'Italie, F-69364 Lyon, France\\[\affilskip]
$^3$ Centre for Stellar and Planetary Astrophysics, School of Mathematical Sciences, Monash University, Clayton Vic 3168, Australia\\[\affilskip]
$^4$ Institut f\"ur Astrophysik der Universit\"at G\"ottingen, Friedrich-Hund-Platz 1, D-37077 G\"ottingen, Germany}

\pubyear{2011}
\volume{123}
\pagerange{111--222}
\date{\today and in revised form ??}

\begin{document}

\maketitle

\begin{abstract}
We present a systematic study of the influence of different forcing types on the statistical properties of supersonic, isothermal turbulence
 in both the Lagrangian and Eulerian frameworks. We analyse a series of high-resolution, hydrodynamical grid simulations with Lagrangian
 tracer particles and examine the effects of solenoidal (divergence-free) and compressive (curl-free) forcing on structure functions,
 their scaling exponents, and the probability density functions of
 the gas density and velocity increments. Compressively driven simulations show a significantly larger density contrast,
 a more intermittent behaviour, and larger fractal dimension of the most dissipative structures at the same root
 mean square Mach number.
 We show that the absolute values of Lagrangian and Eulerian structure functions of all orders in the integral range are only a function of the root
 mean square Mach number, but independent of the forcing.
 With the assumption of a Gaussian distribution for the probability density function of the velocity increments on large scales, we derive a model that describes this behaviour.
\end{abstract}

\section{Introduction}
Knowledge of the statistical characteristics of turbulence is a key prerequisite for understanding turbulent flows on virtually all scales \citep{Frisch1995, Lesieur}.
 While common terrestrial flows are incompressible, astrophysical flows are highly supersonic and compressible. 
 For example, the birth of stars in the interstellar medium is thought to be controlled by
 supersonic turbulence \citep{MacLow2004,ScaloElmegreen2004,McKee2007}.
 As turbulence is by definition a process characterised by a chaotic and irregularly fluctuating velocity field, there is a scale-dependent
 spatial and temporal correlation of fluid quantities \citep{Ishihira2009}.
 The  scale-dependent energy, density and velocity distributions, and the dynamical evolution thereof, are strongly influenced by the properties of the turbulence.
 While large improvements were made in the understanding of incompressible turbulence in the last few years
 \citep[e.g. with new techniques like superstatistics and Lagrangian statistics,][]{Beck2004,Toschi2009},
 there are still open questions in our understanding of compressible turbulence.
 The non-local, inter-scale processes of compressible turbulence arising for example in shock fronts change the Richardson-Kolmogorov
 picture of the energy cascade of incompressible turbulence, where scale-locality is crucial for the existence of universal statistics in the inertial range.
 Basic questions as the existence of the inertial range, the associated scaling laws, and the influence of intermittency are still open.
 Guided by the idea of basic physical quantities like the momentum and the kinetic energy, there is a trend using mass-weighted velocity increments to 
 describe turbulence in a compressible medium. For example, \citet{Kritsuk2007}, \citet{Schmidt2008}, and \citet{Galtier2011} describe the
 intermittency appearing in their supersonic, compressible, numerical simulations with a phenomenological model using
 mass-weighted velocity increments.   
 Since Lagrangian tracer particles are designed to follow the turbulent flow and thus the mass flux in numerical simulations,
 they are a powerful tool to study compressible turbulence, with an intrinsic mass-weighting. Furthermore,
 phenomenological models treating intermittency arising in compressible turbulence were
 developed in the last years. \citet{Boldyrev2002} used the multifractal phenomenological model of \citet{she94}, which describes
 the most dissipative structures in incompressible turbulence as filaments, and modified it such that the most dissipative structures are sheets instead of filaments.
 Nevertheless, despite of these improvements over the last years, 
 the theoretical understanding of compressible turbulence is still poor and analytical derivations are rare.\\
Three spatial or temporal ranges have to be
 distinguished for turbulence: the viscous dissipation range at small scales, the
 inertial range at intermediate scales, and the range above the turbulent injection scale, the so-called integral range.
 The velocity structure functions, an average of the difference of two velocities separated by an increment in space or time, increase from the dissipation range to the integral range.
 Here, we also consider the statistics of turbulence in the integral range, where temporal correlations of the velocity field are exponentially
 damped, and the structure functions saturate over several integral time scales.
 A complete understanding of the statistical properties in this range is necessary as a solid foundation for further studies of the complex and complicated
 behaviour of turbulence in the inertial range. 
 In this paper we shed some light on this unattended topic.
 We compute Lagrangian statistics of the density and
 velocity fields of supersonic turbulence, using data from
 high-resolution, three-dimensional simulations of driven turbulence with
 up to $1024^3$ grid cells and up to $512^3$ tracer particles.
 By calculating probability density functions and structure
 functions, we analyse the sensitivity of compressible turbulence on the forcing of the turbulence.
 In particular, we compare two limiting cases of purely solenoidal (divergence-free) and purely compressive (curl-free) forcing.
 We present a detailed analysis of the scaling properties of the velocity structure functions and intermittency in the inertial subrange.
 We show that a simple analytic formula describes the saturated structure
 functions of all orders in the integral range, with the r.m.s. Mach
 number as single input parameter. This formula can be used to check
 the convergence of the structure functions.\\
In \S~\ref{sec:simulation}, we explain the numerical setup, describe the implementation of the different forcings, the tracer particles, and define the structure functions
 and the statistical moments used to analyse the simulations.
 We analyse structure functions, their scaling properties, intermittency, and the probability density function (PDF) of the mass density and the velocity increments in \S~\ref{sec:results}.
 In \S~\ref{sec:statisticaltheory}, we present a simple analytic formula describing the behaviour of the saturated structure functions of all orders in the integral range.
 A summary of our results and conclusions are given in \S~\ref{sec:conclusion}. 
\section{Simulation and Methods}
\label{sec:simulation}
We solve the hydrodynamical equations on a uniform grid with $256^3$,
 $512^3$, and $1024^3$ grid points, using the piecewise parabolic method
 \citep[PPM, ][]{ColellaWoodward1984}, implemented in the grid code FLASH3 \citep{Fryxell2000, Dubey2008}.
 We start with gas of uniform density at rest and uniformly distributed tracer particles also at rest.
 We place one tracer particle in every other grid cell, such that the simulations contain $128^3$,
 $256^3$, and $512^3$ tracer particles, respectively.
 Since we assume isothermal gas, the pressure, $P= \rho {c_\mathrm{s}}^2$, is proportional to the density $\rho$ with the fixed sound speed $c_\mathrm{s}$.
 We solved the continuity equation and the Euler equation with a stochastic forcing term $\textbf F$,
\begin{equation}
\frac{\partial s}{\partial t} +(\textbf{v} \bcdot \bnabla)s=-\bnabla \bcdot \textbf{v} \,,
\label{eq:kontinuitaets}
\end{equation}
\begin{equation}
\frac{\partial \textbf{v}}{\partial t} +(\textbf{v} \bcdot \bnabla)\textbf{v}=-c_\mathrm{s}^2 \bnabla s +\textbf{F} \, ,
\label{eq:euler}
\end{equation}
where $\textbf{v}$  is the velocity field and $ s \equiv \ln(\rho / \langle\rho\rangle_V)$ is the natural logarithm of the mass
 density divided by the mean (volume-weighted) mass density. 
 The parameters of the simulation are: $\langle \rho \rangle_V =1$,
 $c_\mathrm{s}=1$ and the computational domain has a box length $L=1$ with periodic boundary conditions.
 The numerical simulation is evolved for ten dynamical time scales $T=L/2V$ where $V$ is the integral velocity
 and the relevant quantities are stored in intervals of $0.01\,T$ for the tracer particles and in intervals
 of $0.1\, T$ for the grid.
 We do not apply physical viscosity, so we have to rely on purely numerical viscosity. This requires resolution studies. Hence we investigate our results
 for different grid resolutions: $256^3$,
 $512^3$, and $1024^3$ (Appendix~\ref{sec:resolutionstudy}). It can be shown that the numerical viscosity of PPM can be used as an implicit way of treating physical viscosity,
 as long as a large-enough scale separation is guaranteed \citep{Benzi2008}.\\
\subsection{Forcing module}
\label{subsection:forcing}
The random forcing term $\textbf{F}$ is derived from a stochastic Ornstein-Uhlenbeck process with finite autocorrelation time scale, $T_{\mathrm{ac}}$
 \citep{Eswaran1988, Schmidt2009, Federrath2010}. It gives a stochastic force field $\textbf{F}$ that varies smoothly in space and time.
 The Ornstein-Uhlenbeck process generates the forcing in Fourier space ($k$-space) by solving a differential equation,
\begin{equation}
 \mathrm{d}\widehat{\textbf{F}}(\textbf{k}, t)=F_0(\textbf{k}, T_{\mathrm{ac}})\mathcal{P}^\zeta(\textbf{k}) \frac{\mathrm{d}\textbf{W}(t)}{T_{\mathrm{ac}}}-\widehat{\textbf{F}}(\textbf{k}, t)\frac{\mathrm{d}t}{T_{\mathrm{ac}}}\,,
\label{eq:Forcing}
\end{equation}
where the $\mathrm{d}\textbf{W}(t)$ is a three-dimensional Gaussian random increment with zero mean and standard deviation $\mathrm{d}t$, generated by a Wiener process.
 $\mathcal{P}^\zeta(\textbf{k})$ is a projection tensor in Fourier space. In index notation, this operator is
\begin{equation}
{\mathcal{P}^\zeta}_{ij}(\textbf{k})=\zeta {\mathcal{P}^\bot}_{ij}(\textbf{k})+(1-\zeta) {\mathcal{P}^\|}_{ij}(\textbf{k})\,,
\end{equation}
where $\mathcal{P}^\bot=\delta_{ij}-k_ik_j/k^2$ and $\mathcal{P}^\|=k_ik_j/k^2$ are the fully solenoidal and compressive projection operators respectively. 
 By setting $\zeta =1$, the forcing field  is purely solenoidal (i.e., $ {\bnabla} \bcdot \textbf{F} = 0 $). In contrast, setting $\zeta =0$, the forcing field is purely compressive
 (i.e., $ {\bnabla} \times \textbf{F} = 0 $).
 The natural mixture of forcing modes is obtained for $\zeta =0.5$ \citep{Federrath2010}, which leads to a velocity
 distribution of $ \langle v^2_{\parallel} \rangle /  \langle v^2_{\mathrm{tot}} \rangle= 1/3 $.
 To investigate the influence of the different forcings, we focus on the limiting cases of purely solenoidal forcing ($\zeta=1$) and purely compressive forcing ($\zeta=0$).\\
The forcing amplitude $F_0(\textbf{k},T)$ is a three-dimensional parabolic function, only containing the large (integral) scales $1 <\mid \textbf k \mid < 3$, peaking at $k=2$,
 which corresponds to half of the box size L/2, as we measure $k$ in units of $2\pi / L$. The amplitude of the forcing is adjusted, so that in both cases the volume-weighted r.m.s Mach number is
 $\mathcal{M}_V \approx 5.5 $, when the state of stationary, fully-developed turbulence is reached.\\
The last term in equation (\ref{eq:Forcing}) is a stochastic diffusion term that ensures exponential decrease of the autocorrelation function of the forcing.
 We set the autocorrelation time $T_{\mathrm{ac}}$ of the forcing equal to the dynamical time scale $T$.\\
The density and velocity statistics of turbulence produced by a mainly compressive force field are investigated in \citet{Schmidt2009}, while a systematic statistical comparison of
 solenoidal and compressive forcing is discussed in \citet{Schmidt2008} and in \citet{Federrath2008b, Federrath2009, Federrath2010}.
\subsection{Tracer particles}
We start with uniformly distributed tracer particles at rest.
 Afterwards they can move freely within the computational domain.
 The velocity and density of the tracer particles are calculated with a cloud-in-cell interpolation from the grid at
 the beginning of each time step. Given the interpolated velocity, the tracer particles are then moved with an Euler method,
 based on the hydrodynamical time step. The tracers thus follow the gas flow in the Lagrangian frame of reference.
 Instead of the linear interpolation of the neighbouring grid points, a second-order (triangular-shaped cloud)
 and third-order (tricubic) space interpolation, as well as a higher-order temporal integration scheme (predictor-corrector type)
 were tested, but they did not lead to statistically significant differences.
 As the tracer particles have no influence on the fluid, they are passive tracers of the fluid motion.
\subsection{Velocity increments and structure functions}
In order to calculate the increments, we use the following definition of the time-dependent, Lagrangian velocity increment
\begin{equation}
 \delta v^m_i(t,\tau)= v^m_i(t + \tau) -v^m_i(t),
 \label{eq:Gesch_increm_Lag}
\end{equation}
where $\tau$ is a temporal increment and  $v^m_i(t)$ is the velocity in spatial direction $i \in \{x,y,z\}$ of the $m$th tracer particle at the time $t$.
 The space-dependent, Eulerian velocity increments are defined as
\begin{equation}
 \delta v^{mn}_i(\textbf{r},\boldsymbol{\ell})= v^m_i(\textbf{r} + \boldsymbol{\ell}) -v^n_i(\textbf{r})
 \label{eq:Gesch_increm_Eu}
\end{equation}
\begin{equation}
  \delta v^{mn}_\|(\textbf{r},\boldsymbol{\ell}) = v^m_\| (\textbf{r} + \boldsymbol{\ell}) -v^n_\| (\textbf{r}) \, ,
\end{equation}
where $\textbf{r}$ is the position of the tracer $n$, $\boldsymbol{\ell}$ is the spatial increment between the tracer particles $m$ and $n$,
 and $v_\parallel = \textbf{v} \cdot \boldsymbol{\hat{\ell}} $ with $\boldsymbol{\hat{\ell}} = \boldsymbol{\ell} /\ell $ being the unit vector in the direction $\boldsymbol{\ell}$.
 The Lagrangian structure function (LSF)
\begin{equation}
LS^p(\tau) = \langle \langle |\delta v^m_x(t, \tau) |^p \rangle_m + \langle |\delta v^m_y(t, \tau)|^p \rangle_m + \langle | \delta v^m_z(t, \tau) |^{p} \rangle_m \rangle_t /3	
\label{eq:LS}
\end{equation}
 is obtained by averaging the velocity increments over the different tracer particles $m$, the three directions of the coordinate system ${x, y, z}$
 and over $t \in [2,10]\,T$. This is reasonable because of the time invariance and isotropy in the state of fully-developed turbulence.\\
In practise, we randomly select $5\times10^6$ tracer particles for  all $801$ time samples with $t\geqslant2\,T$ in the  fully-developed state
 for the averaging procedure (\ref{eq:LS}).
 We checked the validity of this approach by doing these calculations also with all $512^3$ tracer particles for one time-line starting at $t=2\,T$ to ensure
 that the used number of sampling pairs has no statistically significant influence on our results (see Appendix~\ref{sec:ConvergenceESF}).
 For calculating the Eulerian structure functions (ESFs)
 \begin{equation}
ES^p(\ell)\equiv \langle \langle | \delta v^{mn}_x(\textbf{r},\ell) |^p \rangle_{mn} + \langle | \delta v^{mn}_y(\textbf{r},\ell)|^p \rangle_{mn} + \langle |\delta v^{mn}_z(\textbf{r},\ell)|^p \rangle_{mn} \rangle_{t} /3 
\label{eq:ES}
\end{equation}
\begin{equation}
ES^p_{\|}(\ell)\equiv  \, \langle | \delta v^{mn}_{\|}(\textbf{r},\ell) |^{p} \rangle_{mn,t}
\label{eq:ES_para}
\end{equation}
the simulation box was divided in $16^3$ sub boxes.
 For $m$, a fixed number of tracer particles is chosen homogeneously distributed over all sub boxes. To obtain a constant sampling of the ESF with $\ell$,
 for each $m$, a subset $\propto 1/{r^2}$  of tracer particles of every sub box is selected for $n$, where $r$ is the distance from $m$ to the centre of the sub box.
 As the number of sub boxes increases proportional to $r^2$, this procedure ensures that for each lag $\ell$, roughly the same number of sampling pairs
 is used. The selection procedure is normalised in a way that for $m$, nearly the same number of tracer particles is selected as for $n$.
 The ESFs are calculated for $81$ snapshots in time intervals of $\Delta t = T/10$ each with about $10^{10}$ sampling pairs.
 We tested our results with different numbers of sub boxes, where with insufficient sub boxes ($\lesssim 8^3$) we have to calculate the ESFs with many more sampling pairs
 to get a good statistic on the smallest scales.
 Using more than $16^3$ sub boxes showed no effective
 improvement of the distribution. With $16^3$ sub boxes, we calculate the ESFs with different numbers of sampling pairs to ensure that increasing the sampling pairs has
 no statistically significant influence on our results.
 We provide detailed convergence tests in Appendix~\ref{sec:ConvergenceESF}.
 Since all increments are calculated on the tracer particles, the structure functions are intrinsically mass-weighted.
\subsection{Statistical Moments}
\label{sec:StatMoments}
In order to calculate the higher-order moments of the PDFs in \S~\ref{sec:results}, we use the following definition for the first four standardised central moments:
\begin{subeqnarray}
 && \textsf{mean:}  \quad \quad\langle q \rangle = \sum q \,p(q)\, \Delta_q \label{eq:mean}\\ 
 && \textsf{standard deviation:}  \quad \quad \sigma_q = \sqrt{\langle (q-\langle q\rangle )^2 \rangle}\\
 && \textsf{skewness:} \quad \quad \mathcal{S}_q = \frac{\langle (q-\langle q\rangle )^3 \rangle}{\sigma^3}\\
 && \textsf{kurtosis:} \quad \quad \mathcal{K}_q = \frac{\langle (q-\langle q\rangle )^4 \rangle}{\sigma^4}\,,
\end{subeqnarray}
where $\Delta_q$ is the bin width of the PDF $p(q)$. With this definition a Gaussian has a skewness $\mathcal{S}_q=0$ and a kurtosis $\mathcal{K}_q=3$. 
\section{Results}
\label{sec:results}
\begin{figure}
\begin{tabular}{cc}
\includegraphics[width=0.5 \linewidth]{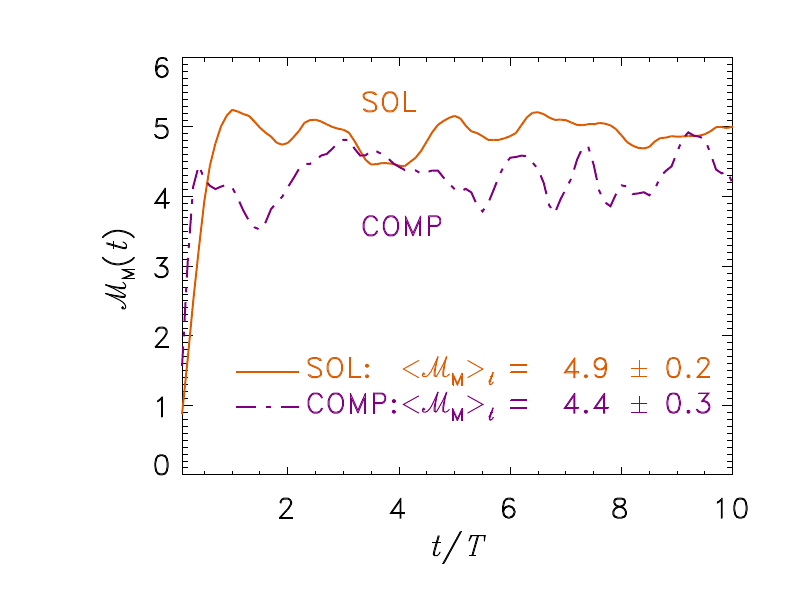}
\includegraphics[width=0.5 \linewidth]{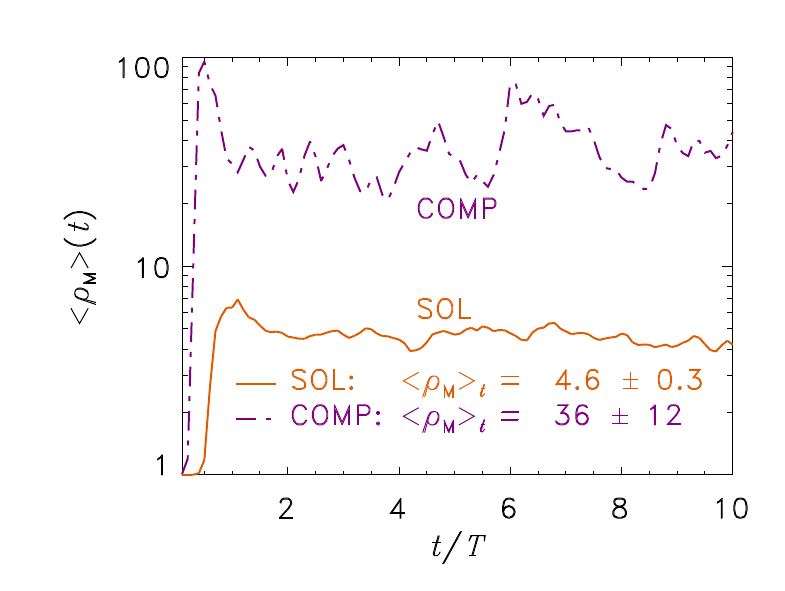}
\end{tabular}
\caption{Mass-weighted r.m.s Mach number (left), and averaged mass-weighted density (right), as a function of the dynamical time, 
 calculated by averaging over all tracer particles.
 Within the first $2\,T$, a statistically steady state was reached for both solenoidal (sol) and compressive (comp) forcing.
 The mean values were averaged over $t \in [2,10]\, T$  and the errors are the $1\,\sigma$ variations in time.}
\label{fig:t_RMS_Dens}
\end{figure}
\begin{figure}
\begin{tabular}{cc}
 \includegraphics[width=0.5 \linewidth]{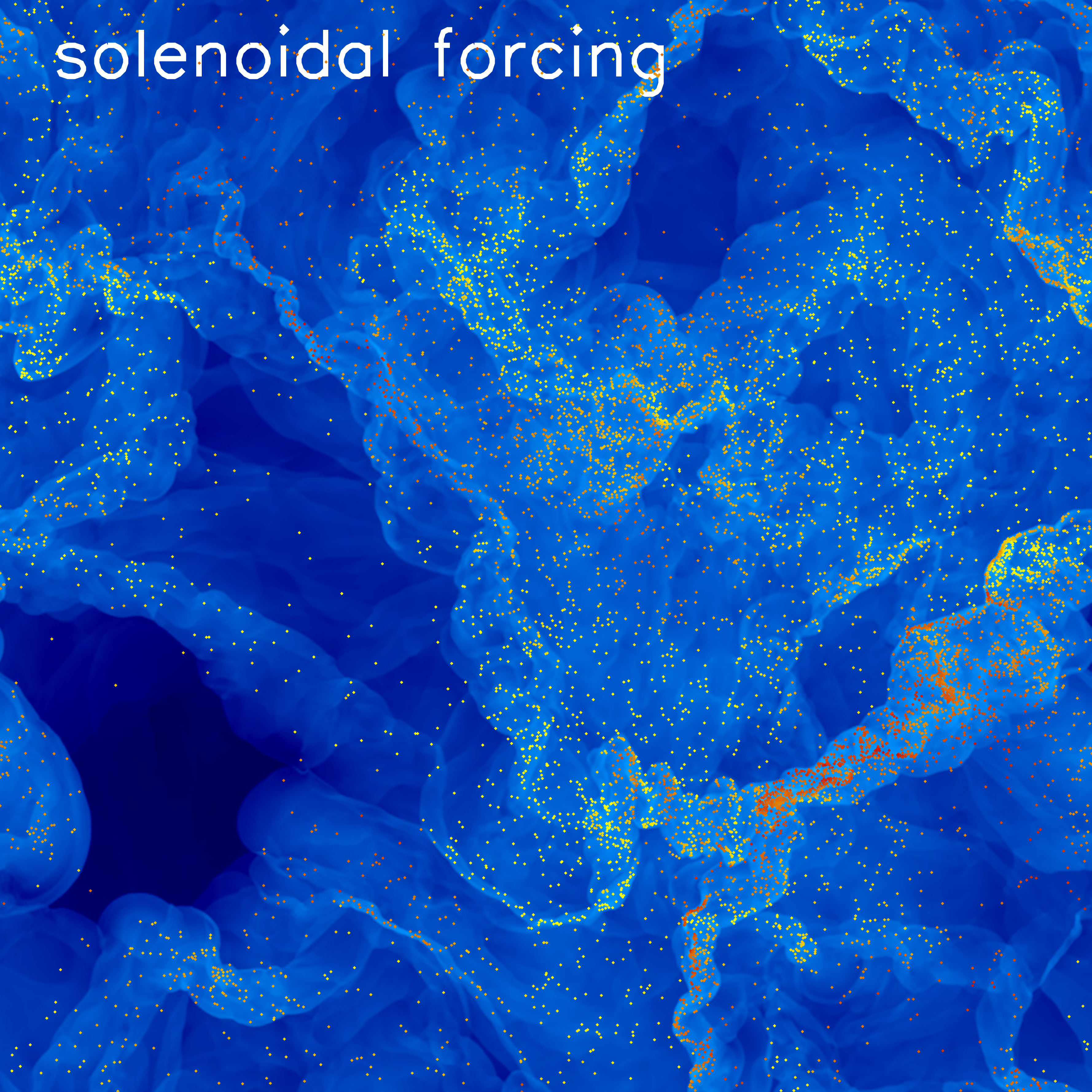}
 \includegraphics[width=0.5 \linewidth]{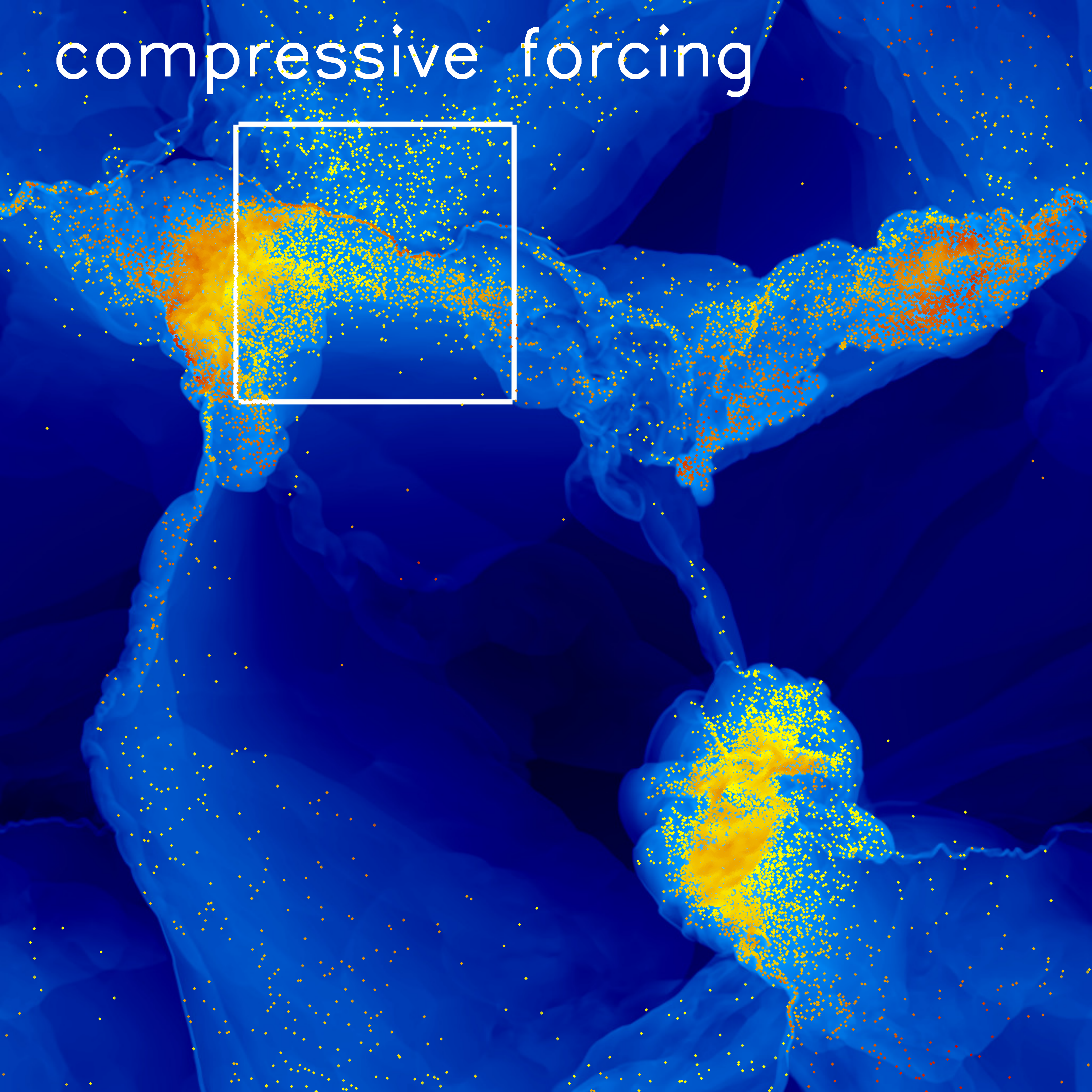}
\end{tabular}
\begin{center}
 \begin{tabular}{cc}
 \includegraphics[width=0.5 \linewidth]{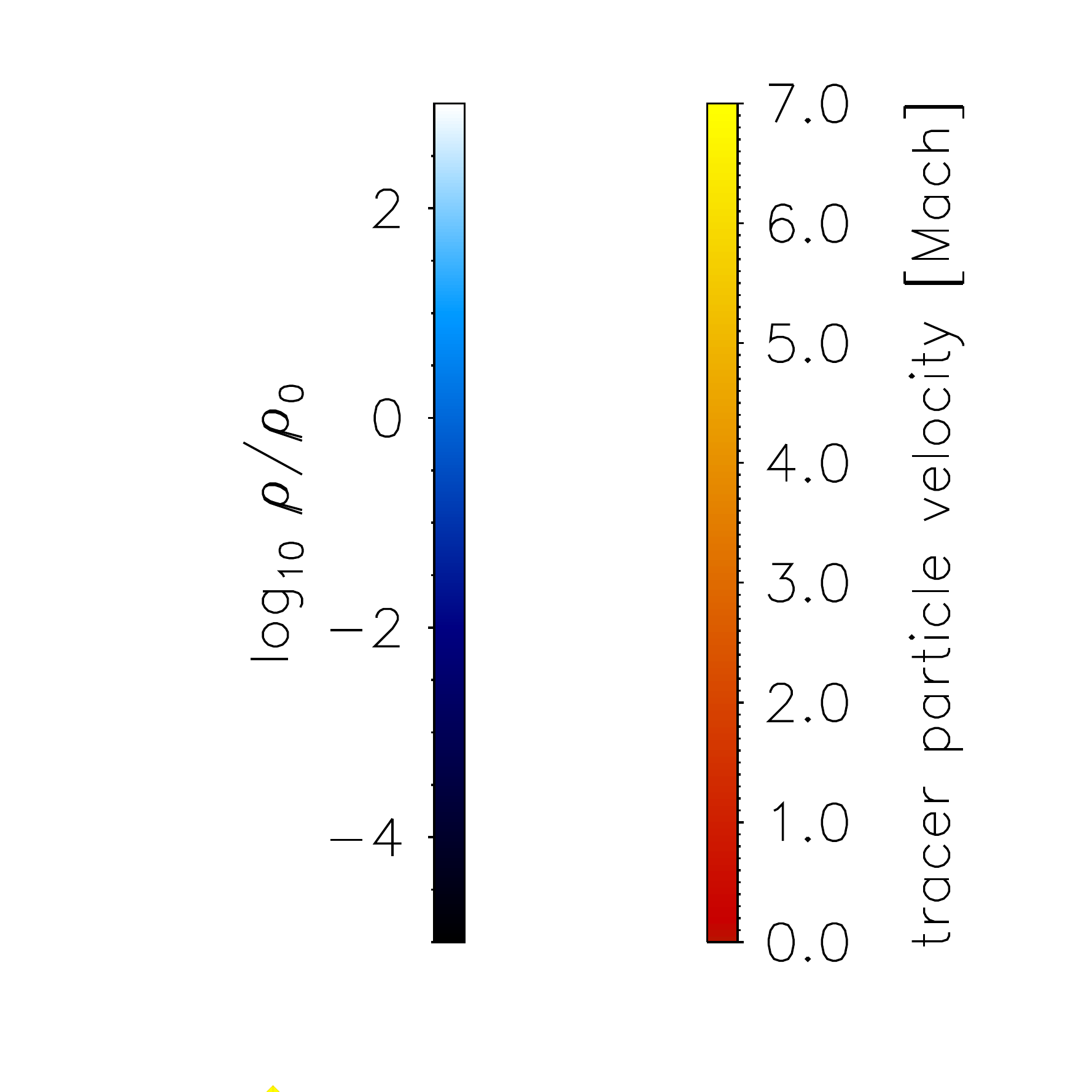}
 \includegraphics[width=0.5 \linewidth]{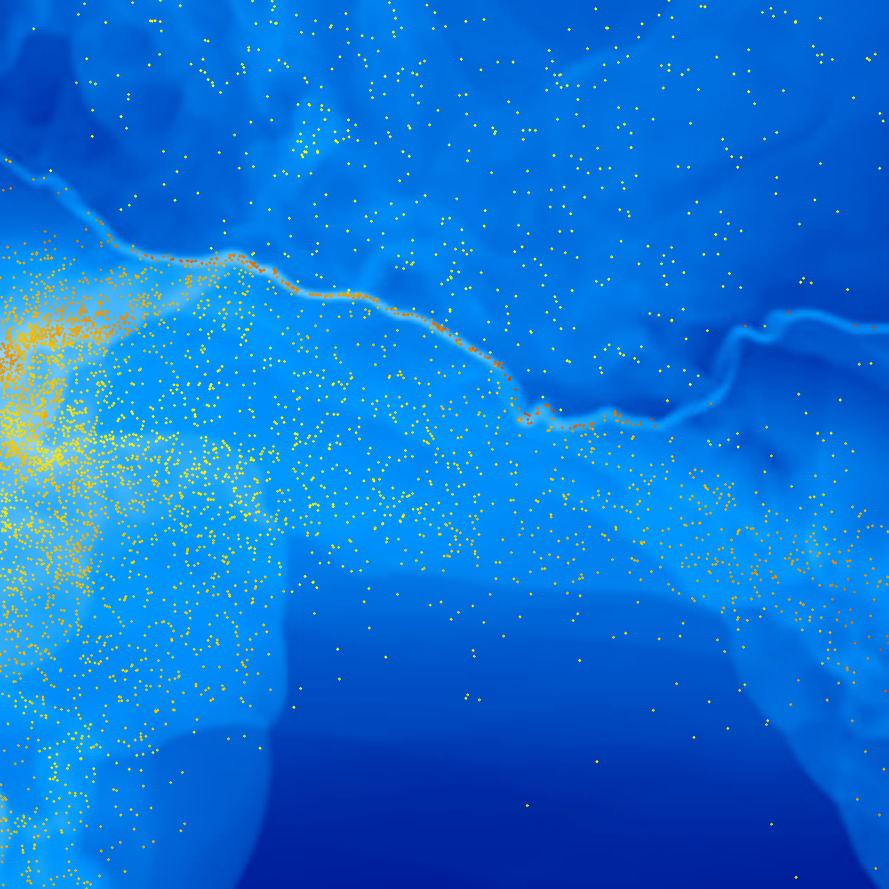}
 \end{tabular}
\end{center}
\caption{Top panels: Slice in the $xy$-plane at $z=0.5$ and time $t=6\,T$ as an example for the regime of statistically fully-developed, compressible turbulence.  
 The logarithm of the mass density of the grid cells in this slice (blue) as well as the norm of the velocity of the tracer particles (red)
 in one slice with thickness of $0.1$ grid cell are displayed for
 solenoidal (left) and compressive forcing (right).
 Bottom right: Magnified slice of a shock front in the simulation with compressive forcing (see white box in upper right panel).
}
\label{fig:DensVel}
\end{figure}	
As discussed in \cite{Federrath2009, Federrath2010} and \cite{Price2010}, the fluid reaches an equilibrium state of fully-developed, supersonic turbulence after about two turbulent crossing times, $t\approx2\,T$.
 We therefore restrict our analysis to times $t\geqslant2\,T$.
 Figure~\ref{fig:t_RMS_Dens} shows the time evolution of the
 mass-weighted r.m.s Mach number and the mass-weighted averaged density, calculated with the data of the tracer particles.
 As the tracer particles are advected by the flow, their density is correlated with the mass density. As a consequence,
 the average of a quantity over all tracer particles is mass-weighted.
 We denote quantities calculated with this quasi Lagrangian statistics with a subscript $M$.
 Figure~\ref{fig:t_RMS_Dens} demonstrates that at $t\approx2\,T$, a regime of statistically fully-developed turbulence is reached.
 The right panel of figure~\ref{fig:t_RMS_Dens} shows that the compressive forcing yields
 a nearly eight times larger mass-weighted mean density and fluctuations thereof.
 The mean values indicated in figure~\ref{fig:t_RMS_Dens} are averaged over the interval $2\,T \leq t \leq 10\,T$, and the errors are the $1\,\sigma$ variations in time.
 In the state of fully-developed turbulence, the time-averaged r.m.s Mach number on the tracer particles is
 $\langle \mathcal{M}_{\mathrm{sol},M} \rangle_t=4.9$ (subscript $t$ for time average) with solenoidal forcing,
 and
 $\langle \mathcal{M}_{\mathrm{comp},M} \rangle_t=4.4$
 with compressive forcing. In \citet[][table~1]{Federrath2008b}, the volume-weighted Mach number of this simulation was measured
 ($\langle \mathcal{M}_{\mathrm{sol},V} \rangle_t=5.3$, $\langle \mathcal{M}_{\mathrm{comp},V}\rangle_t=5.6$).
 Clearly, the intrinsic mass-weighting of the tracer particles influences the statistical properties, discussed in more detail below.
 The values of the volume-weighted and mass-weighted r.m.s. Mach number
 for the different forcings are summarised in table~\ref{tab:Densitymoments} (first row).
 To illustrate the different flow patterns for solenoidal and compressive forcing, we show a randomly selected slice
 through the mid plane of the computational domain at $t=6\,T$ in figure~\ref{fig:DensVel}.
 It shows the logarithm of the mass density computed on the grid cells in this slice,
 as well as the norm of the velocity of the tracer particles that are in one slice with thickness of $0.1$ grid cells.
 Each dot represents a tracer particle with the colour corresponding to the norm of the velocity.
 The density fluctuations are more space-filling with solenoidal forcing, and have smaller amplitude, while compressive forcing yields larger voids and denser regions.
 The bottom panel of figure~\ref{fig:DensVel} shows a magnification of a head-on collision of two flows that leads to a strong shock in the simulation with compressive forcing.
 The density field shows a sharp, well-defined shock front.
 In these compressed regions, the tracer particles accumulate and have a significantly lower Mach number, $\mathcal{M}_M \approx 1-2$.
 These stagnation points, showing a correlation of high density with low velocities \citep[e.g.]{Ballesteros2003, Klessen2005}, are important in the theory of star formation \citep{MacLow2004},
 as they are good candidates for the formation of dense-cores, which are the progenitors of individual stars and binary stellar systems. 
 This correlation in the stagnation points causes the mass-weighted values of the Mach number to be smaller than the volume-weighted ones.
 Compressive forcing excites more head-on collisions and shock fronts, so this effect has a stronger influence in that case. 
\subsection{The probability density function of the gas density}
\label{subs:Dens_PDF}
\begin{figure}
\begin{center}
\begin{tabular}{cc}
\includegraphics[width=0.5 \linewidth]{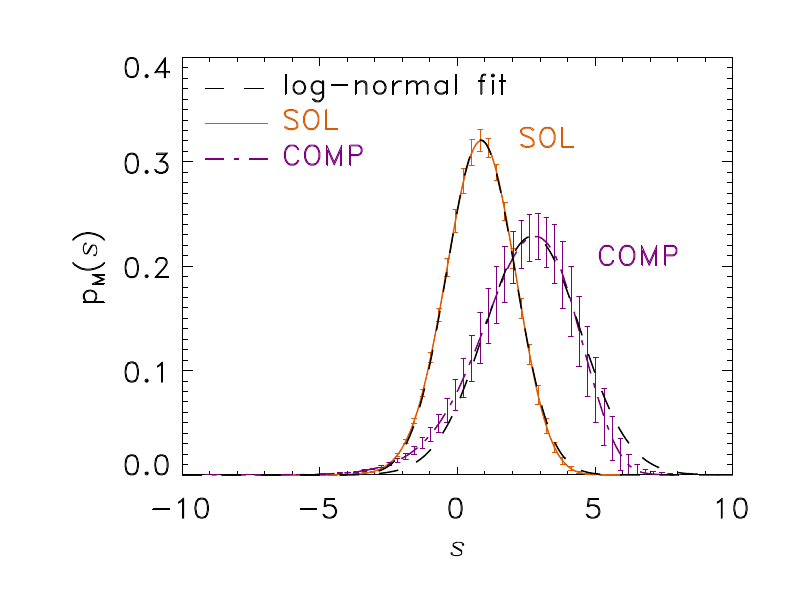}
\includegraphics[width=0.5 \linewidth]{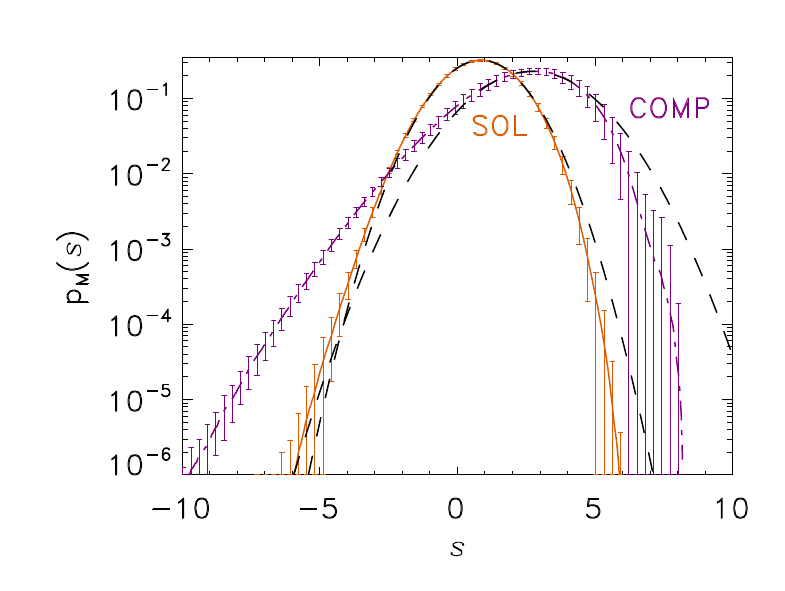}
\end{tabular}
\end{center}
\caption{Mass-weighted PDF $p_M(s)$ of the logarithmic mass density $s = \ln(\rho /\langle \rho \rangle_V)$ for solenoidal (solid line) and compressive (dashed-dotted line) forcing
 in linear (left panel) and logarithmic scaling (right panel), calculated on the tracer particles.
 The PDFs are calculated for $81$ time steps in the state of fully-developed turbulence $t \geqslant 2 T$ and averaged. The error bars indicate the standard deviation of
 the temporal fluctuations. The dashed lines show log-normal fits with the mean value and the standard deviation as fitting parameters.}
\label{fig:Dens_Velocity_PDF}
\end{figure}

The probability density function (PDF) of the gas density $p(\rho)$  and its standard deviation $\sigma_{\rho}$ are important quantities in astrophysics.
 For instance, \cite{Padoan2002} and \cite{Hennebelle2008,Hennebelle2009}
 relate the density PDF to the mass distribution of dense gas cores and stars.
 \citet{Padoan1997} and \citet{Passot1998} have shown that the standard deviation grows proportional to the Mach number of the turbulent flow, if the density PDF is
 close to a log-normal distribution \citep[see][for a recent, extended study]{Price2011}. \cite{Federrath2010} demonstrated that the density PDF is not only influenced by the r.m.s Mach number,
 but also by the forcing parameter $\zeta$, and presented a modification of the existing expression, which takes the ratio of solenoidal and compressive modes of the forcing into account.
 In many numerical experiments of driven, supersonic, isothermal turbulence with solenoidal and$/$or weakly compressive forcing, it was found that the density PDF
 is close to a log-normal distribution \citep[e.g. ,][]{Padoan1997, Klessen2000, Lemaster2008,Federrath2008b},
\begin{equation}
 p(s)=\frac{1}{\sqrt{2\pi}\sigma_{s}}\exp{\left(\frac{-(s-\langle s \rangle)^2}{2\sigma_s^2}\right)}\,, 
\end{equation}
where $s=\ln(\rho / \langle\rho\rangle_V)$ is the logarithm of the density divided by the volume-weighted mean density.
\cite{Li2003} argued that 
 the mass-weighted density distribution is also a log-normal, with the same standard deviation as the volume-weighted distribution.
 With the assumption of a log-normal density PDF, the authors derived a relation between the mass-weighted and the volume-weighted quantities, 
\begin{equation}
 \langle s \rangle_{V}=-\langle s \rangle_{M} = -\frac{\sigma_s^2}{2} \, .
\label{eq:Lee}
\end{equation}
Figure~\ref{fig:Dens_Velocity_PDF} presents the time-averaged ($t \in [2,10]\,T$) PDF of the quantity $s_M$, calculated on the tracer particles 
 in linear (left panel) and logarithmic scaling (right panel).
 The first four standardised central moments (see section~\ref{sec:StatMoments}) of the density PDF $p_M(s)$ are summarised in table~\ref{tab:Densitymoments} together with the volume-weighted moments
 calculated in \citet[][table~1]{Federrath2010}.
 Compressive forcing yields a significantly broader mass-weighted density distribution with
 standard deviation, $\sigma_{s,M}$ about $1.5$ times larger at the same volume-weighted Mach number than solenoidal forcing ($\sigma_{s, \mathrm{sol}, M}=1.23$, $\sigma_{s, \mathrm{comp},M}=1.77$).
 The peak is shifted to larger values of the logarithmic density ($\langle s\rangle_{\mathrm{sol},M}=0.81$, $\langle s\rangle_{\mathrm{comp},M}=2.37$).
 The PDF is compatible with a Gaussian distribution for solenoidal forcing. However, the Gaussian fit (black dashed line) shows that
 the density PDF has weak non-Gaussian contributions in the wings of the distribution.
 On the other hand, in the density PDF obtained from compressive forcing, the discrepancy to the Gaussian distribution in both wings is more prominent.
 The deviations from the log-normal distribution for the compressive forcing is caused by both a physical and a numerical effect.
 \citet{Price2010} analysed the influence of measuring the density PDF by using a grid-based simulation and an SPH simulation and found
 that the PDF of the SPH particles increases slightly in the high-density tail with increasing resolution and decreases in the low-density tail.
 We expect that this effect will decrease the deviations from a log-normal distribution for both forcing types in our simulation as the resolution is increased.
 \citet[][figure~6]{Federrath2010} analysed the volume-weighted density PDFs for resolutions of $256^3$, $512^3$, and $1024^3$, showing that changing the resolution affects solenoidal
 and compressive forcing in roughly the same way. \citet{Schmidt2009} also argue that the deviations from a log-normal PDF produced by compressive forcing are a genuine effect.
 From this we can conclude that the stronger non log-normal features seen for compressive forcing likely have a physical origin rather than a purely numerical one.
 The higher moments of the distribution with compressive forcing ($\mathcal{S}_{\mathrm{comp},M} = -0.57$,
 $\mathcal{K}_{\mathrm{comp},M}=3.50$) show larger deviations from the Gaussian values ($\mathcal{S} = 0$, $\mathcal{K}=3$) than
 for solenoidal forcing ($\mathcal{S}_{\mathrm{sol},M} = -0.13$, $\mathcal{K}_{\mathrm{sol},M}=2.95$).
 Checking the relation (\ref{eq:Lee}) between the mean value and the standard deviation demonstrates that for solenoidal forcing, the 
 assumption of a log-normal PDF is nearly fulfilled ($\sigma_{s, \mathrm{sol},M}^2/2=0.76$).
 In contrast, we find larger discrepancy for compressive forcing ($\sigma_{s,\mathrm{comp},M}^2/2 = 1.57$).
 Measuring the volume-weighted PDF, \cite{Federrath2010} also reported small deviations from a Gaussian distribution for solenoidal forcing ($\mathcal{S}_{\mathrm{sol},V} = -0.10$, $\mathcal{K}_{\mathrm{sol},V}=3.03$)
 and larger deviations for compressively-driven turbulence ($\mathcal{S}_{\mathrm{comp},V} = -0.26$, $\mathcal{K}_{\mathrm{comp},V}=2.91$).
 The mass-weighted quantities show larger discrepancy from the Gaussian values than the volume-weighted quantities.\\
\citet{Padoan1997} and \citet{Passot1998} motivated a linear relation between the r.m.s Mach number and
 the standard deviation of the linear density,
\begin{equation}
 \sigma_{\rho}/\rho_0 = b\, \mathcal{M}\,,
 \label{eq:b}
\end{equation}
with a proportionality constant $b$.
 Several measurements of $b$ were obtained from different simulations and with different types of forcing,
 and yielded $b$ between $0.26$ and $1.05$ \citep{Passot1998, Li2003, Kritsuk2007,Beetz2008, Federrath2008b, Schmidt2009, Price2011}.
 From the distribution $p_M(s)$ shown in figure~\ref{fig:Dens_Velocity_PDF}, we calculate $p_M(\rho)$.
 Since equation (\ref{eq:b}) was derived with volume-weighted quantities, we have to transform our results using the relation \citep[see,][]{Li2003},
\begin{equation}
p_M(\rho)\propto \frac{\mathrm{d}M}{\mathrm{d}V} \frac{\mathrm{d}V}{\mathrm{d}\rho}\propto \rho\,p_V(\rho)\,,
\end{equation}
where  $p_M(\rho)$ and $p_V(\rho)$ are the mass-weighted and the volume-weighted PDFs of the gas density, respectively.
 We find $\sigma_{\rho,\mathrm{sol},V} = 1.90$ for purely solenoidal and $\sigma_{\rho,\mathrm{comp},V} = 6.03$ for purely compressive forcing.
 With the volume-weighted Mach number of this simulation, we get $b_{\mathrm{sol}}=0.36$ for solenoidal forcing and $b_{\mathrm{comp}}=1.08$ for compressive forcing,
 in good agreement with \cite{Federrath2008b}. 
\begin{table}
\begin{center}
\begin{minipage}{12cm}
\begin{tabular}{c|cc|cc}
				   &	\multicolumn{2}{c}{solenoidal}			       &\multicolumn{2}{c}{compressive}		\\
{} 				   &  {mass-weighted }        & {volume-weighted}              &{mass-weighted}         &{volume-weighted}\\[3pt]
$\mathcal{M}$	                   & $4.9 \pm 0.2$ 	      & $5.3 \pm 0.2$	       & $4.4 \pm 0.3$          & $5.6 \pm 0.3$\\
$\langle s \rangle$                & $0.81 \pm 0.04$  	      & $-0.83 \pm 0.05$       & $2.37 \pm 0.24$ 	& $-3.40 \pm 0.43$\\
$\sigma_s$           		   & $1.23 \pm 0.03 $         & $1.32 \pm 0.06$        & $1.77 \pm 0.09$       	& $ 3.04 \pm 0.24$\\
$\mathcal{S}_s$           	   & $-0.13 \pm 0.05$         & $-0.10 \pm 0.11$       & $-0.57 \pm 0.17$      	& $-0.26 \pm 0.20$\\
$\mathcal{K}_s$    	   	   & $2.95 \pm 0.07$ 	      & $3.03 \pm 0.17$	       & $3.50 \pm 0.33$	& $ 2.91 \pm 0.43$\\
\end{tabular}
\end{minipage}
\end{center}
\caption{Mass-weighted and volume-weighted r.m.s. Mach number (first row) and the first four standardised central moments of the PDF $p_M(s)$ (column 1 and 3)
 and $p_V(s)$ (column 2 and 4) for solenoidal and compressive forcing.
 These quantities are averaged in time when the equilibrium state of fully-developed, supersonic turbulence is reached (see \S\ref{sec:results} and \S\ref{subs:Dens_PDF})
 and the errors are the standard deviation in time.
}
 \label{tab:Densitymoments}
\end{table}
\subsection{The probability density functions of velocity increments}
\begin{figure}
\begin{center}
\includegraphics[width=1 \linewidth]{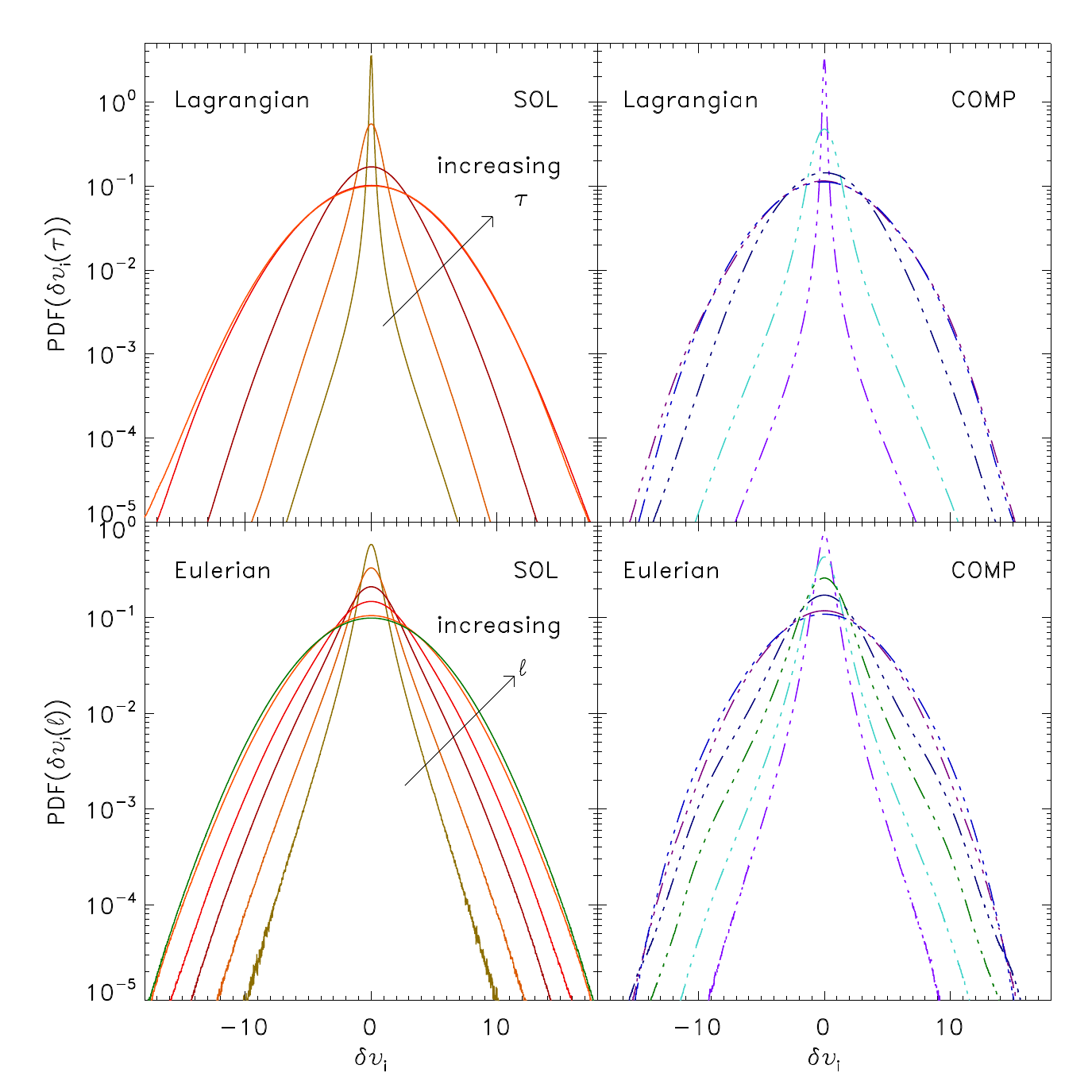}
\end{center}
\caption{Top panel: PDFs of the velocity increment $\delta v_i$ in the Lagrangian framework for solenoidal (left panel) and  compressive forcing (right panel). It shows the
 PDFs with five different temporal increments $\tau \in \{ 0.01 ,\,0.08 ,\, 0.4 ,\,2 ,\,4 \}T$.
 Bottom panel: PDFs of the velocity increment $\delta v_i$ in the Eulerian framework for solenoidal (left panel) and  compressive forcing (right panel). It shows
 the PDFs with six different spatial increments $\ell \in \{ 0.006 ,\,0.02 ,\,0.06,\,0.12,\, 0.25,\,0.49 \}L$.
 For small temporal or spatial lags, $\tau$ and $\ell$, respectively, the PDFs differ from a Gaussian distribution because of intermittency.
 For large $\tau$ or $\ell$, they converge towards a Gaussian distribution.}
\label{fig:PDF_VELInkrements}
\end{figure}

\begin{figure}
\begin{center}
\includegraphics[width=1 \linewidth]{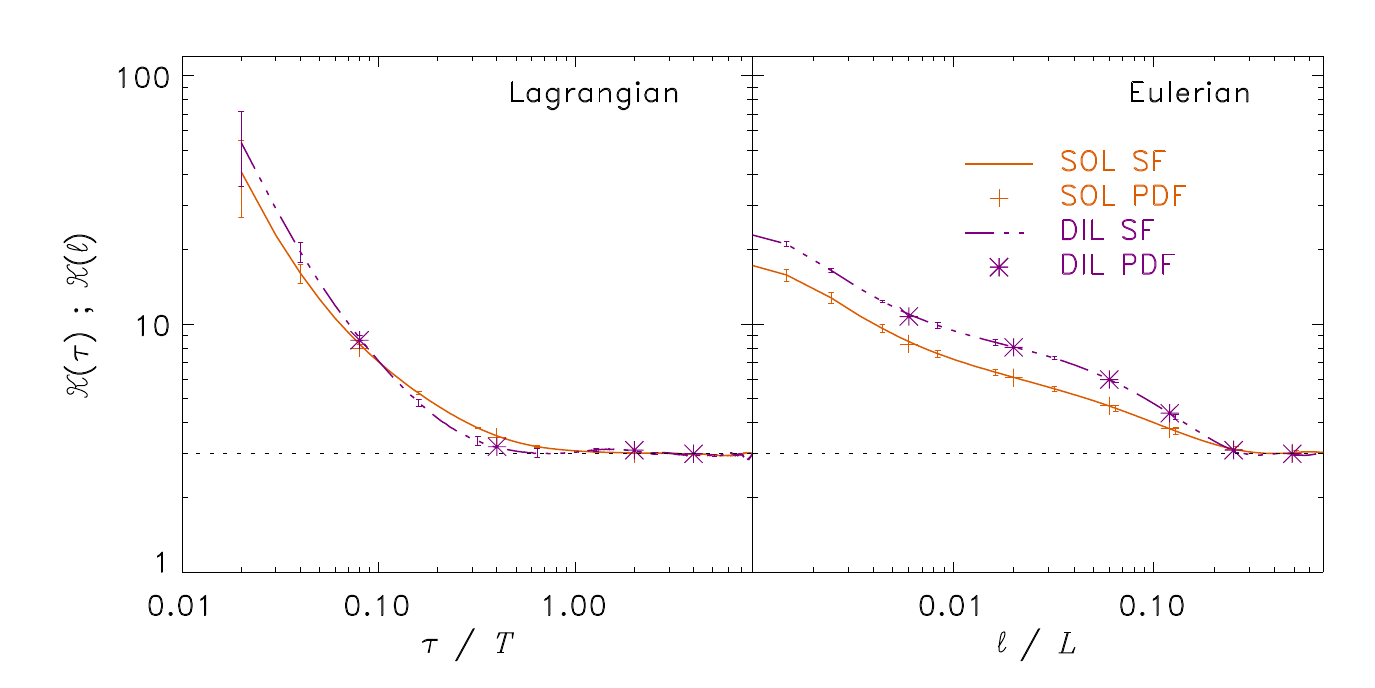}
\end{center}
\caption{Kurtosis as a function of the temporal or spatial increment $\tau$ (left) and $\ell$ (right). The solid and dashed-dotted lines are respectively,
 the values for solenoidal and compressive forcing, calculated
 from (\ref{eq:Kurtosis_SF}). The crosses and stars are respectively the values for solenoidal and compressive forcing, calculated
 with the five PDFs shown in figure~\ref{fig:PDF_VELInkrements}.
 The horizontal, dotted line is the value of a Gaussian distribution.}
\label{fig:Kurtosisverlauf}
\end{figure}

The simplest set of correlation functions to quantify the statistical properties in a compressible, supersonic turbulent flow is the distribution of the velocity increments
 and its higher moments, the structure functions, defined by (\ref{eq:LS} to \ref{eq:ES_para}).
 The deviation of the structure function scaling exponents from the predicted values of the Kolmogorov model \citep{Kolmogorov41} is an effect of intermittency \citep[e.g.,][]{she94}.
 A property of intermittency is that the PDFs of the velocity fluctuations become more and more non-Gaussian on smaller and smaller scales \citep{Gotoh2002,Mordant2002}.\\
Figure~\ref{fig:PDF_VELInkrements} shows the PDF of the velocity increment $\delta v_i$ in the Lagrangian framework for five
 temporal increments $\tau \in \{0.01,\,0.08 ,\, 0.4 ,\,2 ,\,4 \}T$
 and in the Eulerian framework for six spatial increments $\ell \in \{ 0.006 ,\,0.02 ,\,0.06,\,0.12,\, 0.25,\,0.49 \}L$.
 The PDFs follow a Gaussian distribution for $\tau \rightarrow T$ and $\ell \rightarrow L$.
 Decreasing the spatial or temporal increment,
 the Gaussian PDFs vary continuously towards distributions with exponential wings indicating the intermittent behaviour of the turbulent velocity field.
 Figure~\ref{fig:Kurtosisverlauf} shows the kurtosis (see \S~\ref{sec:StatMoments}) of the distributions of the Lagrangian (left panel) and Eulerian (right panel) velocity increments,
 calculated with the structure functions 
\begin{equation}
\mathcal{K}(\tau)=LS^4(\tau)/[LS^2(\tau)]^2 
\label{eq:Kurtosis_SF}
\end{equation}
(solid and dash-dotted lines) and the values calculated with the PDFs (crosses and stars).
 The kurtosis can be used as a measure for the deviations of the distributions of the velocity increments from a Gaussian distribution.
 In the Lagrangian framework, the kurtosis obtained with solenoidal forcing converges towards the Gaussian value ($\mathcal{K}=3$)
 on times comparable with the dynamical time scale, $\tau \approx 1\,T$.
 The compressive forcing yields PDFs converging already on smaller temporal lags, $\tau \approx 0.7\,T $ than solenoidal forcing.
 Compressive forcing develops a more intermittent behaviour with larger kurtosis than the solenoidal forcing for times $\tau \lesssim 0.08\,T $.
 As the non-Gaussian wings of PDFs of the density and/or velocity is caused by intermittency \citep[see][]{Federrath2010},
 this analysis of the kurtosis and the more intermittent behaviour of the compressive forcing
 confirm our observation of the density PDF and its deviation from the log-normal distribution (see \S~\ref{subs:Dens_PDF}).\\
In the Eulerian framework, compressive forcing yields a more intermittent behaviour with a larger kurtosis on nearly all spatial scales than solenoidal forcing.
 However, the kurtosis obtained with compressive forcing converges at the same scale ($\ell \approx 0.23\,L$) towards the Gaussian value as
 the kurtosis of the solenoidal forcing.
 Comparing the kurtosis of the Lagrangian and Eulerian structure functions strengthens the conclusion that Lagrangian statistics
 are more intermittent than Eulerian ones (here shown for two limiting types of forcing)
 as already observed by \cite{Benzi2010}, but only for purely solenoidal forcing.
 In the region, where the kurtosis is converged towards the Gaussian value
 ($\tau \geqslant2.5\,T$ for the Lagrangian framework and $\ell \geqslant 0.4\,L$ for the Eulerian framework), we average
 the structure functions to calculate the mean value
 of the saturated structure functions, as discussed in \S~\ref{subs:LSF_ESF}.
\subsection{Lagrangian and Eulerian structure functions}
\label{subs:LSF_ESF}
\begin{figure}
\begin{center}
\includegraphics[width=1 \linewidth]{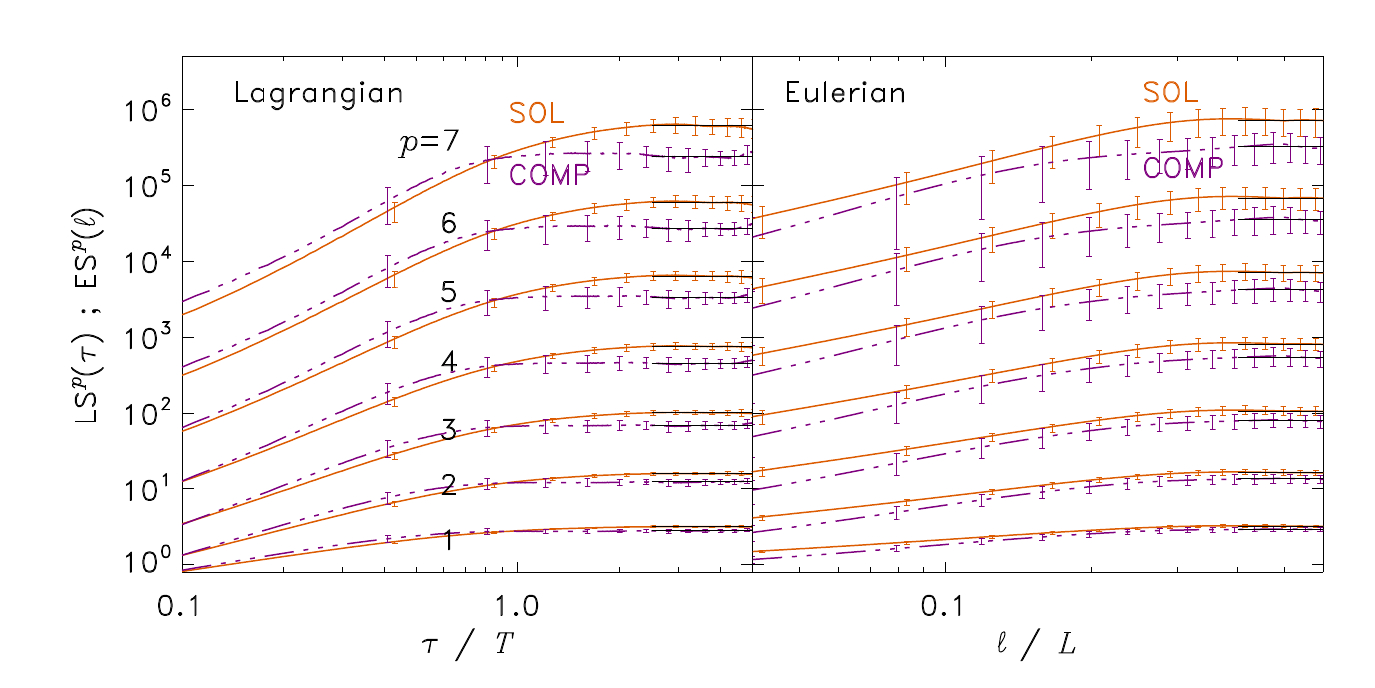}
\end{center}
\caption{
Lagrangian (left) and Eulerian (right) structure functions up to the order of $p=7$ for solenoidal and compressive forcing,
 calculated with the velocity increment of the tracer particles. The error bars of the structure functions indicate the standard deviation in time.
 To calculate the absolute values of the saturated structure functions,
 the Lagrangian structure functions
 were averaged in the range $\tau \in [2.5,5]\,T$, and the Eulerian structure functions
 were averaged in the range $\ell \in [0.4,0.7]\,L$, indicated by the horizontal black lines in the integral range.
}
\label{fig:Strukturfunktionen}
\end{figure}

\begin{figure}
\begin{center}
\includegraphics[width=1 \linewidth]{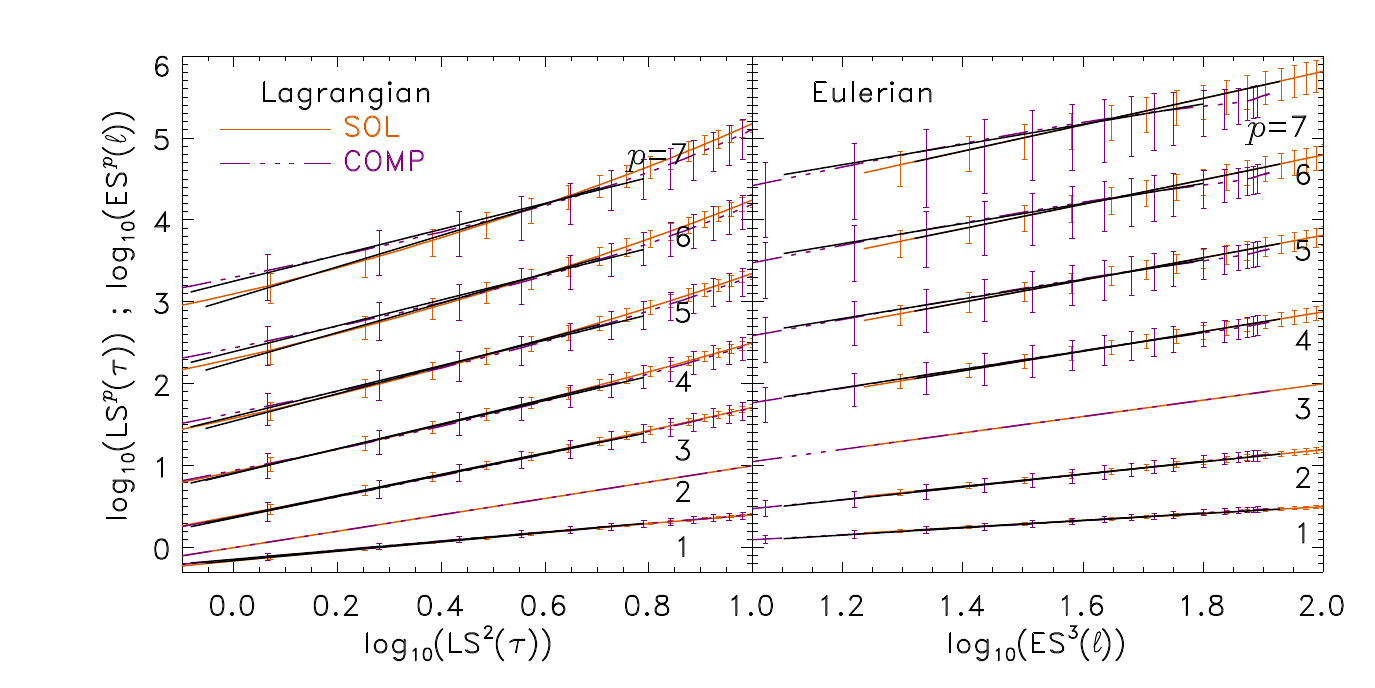}
\end{center}
\caption{Extended self-similarity for the Lagrangian (left panel) and Eulerian (right panel) structure function for solenoidal and compressive forcing. The black lines indicate a linear fit
 for the calculation of the relative scaling exponents, which are summarised in table~\protect\ref{tab:ScalingExponents}.}
\label{fig:ESS}
\end{figure}

Figure~\ref{fig:Strukturfunktionen} shows the LSF and ESF up to order $p=7$ for solenoidal and compressive forcing.
 We calculate the saturation values of the structure functions on the largest scales by averaging them in the range $\tau \in [2.5,5]\,T$ for the LSFs
 and $\ell \in [0.4,0.7]\,L$ for the ESFs.
 The result is displayed as black lines in figure~\ref{fig:Strukturfunktionen}.
 The compressive forcing yields structure functions that converge to Gaussian values already on smaller scales, as observed in figure~\ref{fig:Kurtosisverlauf}.
 For the two forcings, the saturation values are different. This can be explained with the different mass-weighted r.m.s Mach number $\mathcal{M}_M$, observed in figure~\ref{fig:t_RMS_Dens},
 because the structure functions $S$ of order $p$ for infinite increments are $S^p(\infty) \propto \mathcal{M}_M^p$
 (see discussion in \S~\ref{sec:statisticaltheory}).\\
 In turbulence theory, the structure functions follow a power law in the inertial range
\begin{equation}
 LS(p) \propto \tau^{\xi(p)},\quad ES(p)\propto \ell^{\zeta(p)}\,,
\end{equation}
with the scaling exponents $ \xi(p)$ and $ \zeta(p)$. To calculate these scaling exponents,
 we use the inertial range as constrained by \citet{Federrath2009, Federrath2010} ($0.067 \lesssim \ell /L \lesssim0.2$), and transform it to the Lagrangian framework
 ($0.16 \lesssim \tau / T \lesssim 0.34 $) with
 $\tau \propto \ell^{2/3}$. This relation follows directly from the Kolmogorov \textit{four-fifth} law \citep[e.g.,][]{Frisch1995}, implying that the third-order
 structure function scales linearly with $\ell$ in the Eulerian framework. In Burgers turbulence, $\tau \propto \ell$, which
 follows from the assumption of a constant averaged energy transport through the scales, $\bar{\epsilon} \propto E(\ell) / \tau \propto v(\ell)^2 / \tau$,
 and assuming Burgers scaling for the second-order velocity increment, $\delta v^2(\ell) \propto v^2(\ell) \propto \ell$. 
 Using Burgers scaling for $\tau$  in the transformation of the inertial range leads to ($0.067 \lesssim  \tau / T \lesssim0.2$) in the Lagrangian framework.
 The so-called method of extended self-similarity (ESS) proposed by \citet{Benzi98}
 allows for an increased scaling range
 between the smallest scales, influenced by the resolution, and the largest scales with a direct
 influence of the forcing. 
 Using ESS, we thus extend the fitting range to $0.067 \lesssim  \tau / T \lesssim0.34$, which covers both the transformation with $\tau \propto\ell^{2/3}$
 and $\tau \propto \ell$. For the Eulerian structure functions, we extended the scaling range to ($0.05 \lesssim \ell /L \lesssim0.22$) for which we obtain
 a reasonable power-law scaling with ESS.
 Figure~\ref{fig:ESS} shows the ESS scaling plots, i.e., plots of the logarithm of the structure functions calculated with equations (\ref{eq:LS}) and (\ref{eq:ES})
 for the different orders as a function of the logarithm of the second- and third-order
 structure function in the Lagrangian and Eulerian framework, respectively.
 The black lines indicate linear fits for the ESS-measurement of the relative scaling exponents,
\begin{equation}
 Z_L(p)=\frac{\xi(p)}{\xi(2)}, \quad Z_E(p)=\frac{\zeta(p)}{\zeta(3)}\,,
\end{equation}
which are summarised in table~\ref{tab:ScalingExponents} for the Lagrangian framework (2th and 3th columns) and the Eulerian framework
 (5th and 6th columns) for solenoidal and compressive forcing, respectively.
\begin{table}
\begin{center}

\begin{minipage}{14cm}
{
\small

\begin{tabular}{l|ccc|ccccc}
		&		\multicolumn{3}{c}{Lagrangian}	&  \multicolumn{5}{c}{Eulerian}       \\
  &	& & & & & & & \\
{$p$} 	        &{$Z_L(p)$}      & {$Z_L(p)$}     &{BBFLT10}      &{$Z_E(p)$}     &{$Z_E(p)$}       &{SFK08}  &{SFK08 } &{GFN02}\\[3pt]
		& {sol.}	 & {comp.}	  &		  &{sol.}	  &{comp.}	    &{sol.}   &{comp.}	&(transversal)     \\
\hline
$1$             & $0.57 \pm 0.02$ & $0.55 \pm 0.04$&		   &$0.422 \pm 0.008$&$0.45 \pm 0.02$ &$ 0.539 $&$ 0.605 $&$0.369 \pm 0.004$\\
$2$             & $1.0          $ & $1.0          $&$1.0$	   &$0.75 \pm 0.02$ &$0.78 \pm 0.03$ &$ 0.840 $&$ 0.869 $&$0.701 \pm 0.01$ \\
$3$		& $1.30 \pm 0.05 $& $1.30 \pm 0.09$&		   &$1.0          $ &$1.0          $ &$ 1.0   $&$ 1.0   $&$0.998 \pm 0.02$ \\
$4$		& $1.51 \pm 0.07 $& $1.5  \pm 0.1 $&$1.66 \pm 0.02$&$1.20 \pm 0.04$ &$1.13 \pm 0.07$ &$ 1.080 $&$ 1.066 $&$1.26  \pm 0.03$ \\
$5$		& $1.67 \pm 0.09 $& $1.6  \pm 0.1 $& 		   &$1.36 \pm 0.05$ &$1.19 \pm 0.09$ &$ 1.112 $&$ 1.100 $&$1.49  \pm 0.04$ \\
$6$		& $1.8  \pm 0.1  $& $1.6  \pm 0.2 $&$2.10 \pm 0.10$&$1.49 \pm 0.07$ &$1.2  \pm 0.1 $ &$ -     $&$ -     $&$1.69  \pm 0.05$ \\
$7$		& $1.9  \pm 0.1  $& $1.6  \pm 0.2 $& 		   &$1.61 \pm 0.09$ &$1.2  \pm 0.2 $ &$ -     $&$ -     $&$1.86  \pm 0.05$ \\
\end{tabular}
}
\end{minipage}

\end{center}
\caption{Relative scaling exponents of the structure functions $LS^p(\tau)$ and $ES^p(\ell)$ calculated with the ESS method \protect\citep{Benzi98}
 for solenoidal and compressive forcing in the Lagrangian (2th and 3th columns)  and Eulerian (5th and 6th columns) frameworks.
 For a comparison with the scaling exponents of the structure functions in an incompressible turbulent medium, we refer to the data
 of numerical simulations published by \protect\citet{Benzi2010} (BBFLT10, 4th column) and \protect\citet{Gotoh2002} (GFN02, 9th column).
 The 7th and 8th columns show the relative scaling exponents of the transversal structure function of our simulation, but calculated with the 
 velocities measured on the grid and mass-weighted with $\rho^{1/3}$ published in \protect\citet{Schmidt2008}(SFK08).
 }
 \label{tab:ScalingExponents}
\end{table}
To compare our results with data from incompressible turbulence, we refer in table~\ref{tab:ScalingExponents}
 to the data of numerical simulations of subsonic turbulence
 published by \citet[][table 2, Reynolds$_{\lambda}\sim600$, BBFLT10]{Benzi2010} and \citet[][table 3, Reynolds$_{\lambda}=381$, GFN02]{Gotoh2002}.
 We only show the data of the transverse structure functions of GFN02 in the Eulerian framework, because the differences between the longitudinal and transverse structure
 function are negligible compared to the differences between the supersonic and the subsonic results. 
 We expect that our results of the structure function, averaged over the three directions of the coordinate system are in-between the results of
 the longitudinal and transverse structure function. Additionally, we compare our results calculated with tracer particles, discussed here, with the results of
 the same simulations, but calculated with the $\rho^{1/3}$ mass-weighted velocities measured on the grid published by \protect\citet{Schmidt2008} (SFK08, 7th and 8th columns).
 These scaling exponents are also calculated for the transverse structure functions.
 The mass-weighting $\rho^{1/3}$, used by many authors \citep[see e.g.,][]{Kritsuk2007, Kowal2007, Schmidt2008, Galtier2011},
 follows from the assumption of a constant mean volume energy transfer rate in a statistically steady state, $\rho v^2 v/\ell$, so that $\rho v^3  \propto\ell$.
 With the data in table~\ref{tab:ScalingExponents} we can analyse the influence of the different forcings in each framework. 
 The relative scaling exponents show only a significant difference between the scaling behaviour of the solenoidal and compressible forcing in the Lagrangian framework for the highest order.
 The scaling exponents of the compressive forcing are slightly below the scaling exponents of the solenoidal forcing for higher orders.
 However, in the Eulerian framework, this effect is stronger and the compressive forcing causes scaling exponents to stay nearly constant above an order $p>4$,
 such that there is a significant difference between the scaling exponents of the solenoidal and compressive forcing.
 With the measured scaling exponents, we can quantify the intermittency in the different frameworks by calculating the differences to
 the predicted \citet[][K41]{Kolmogorov41} scaling. In the Lagrangian framework and for the highest order, $p=7$, the scaling exponents are $46 \pm 3\%$ and $54 \pm 6\%$
 smaller than the K41 value
 for the solenoidal and compressive forcing, respectively. In the Eulerian framework, the scaling exponents are $31 \pm 4\%$ and $49 \pm 9\%$ smaller than the K41 value.
 For solenoidal forcing, the scaling exponents show a more intermittent behaviour in the Lagrangian framework than in the Eulerian one.
 This is consistent with our analysis of the kurtosis in figure~\ref{fig:Kurtosisverlauf}
 and the results of \citet{Benzi2010}. For compressive forcing, we have an intermittency of the same order for both frameworks.
 The intense density fluctuations in the simulation with compressive forcing cause a more intermittent behaviour and scaling exponents that deviate stronger from the K41 values.
 The stronger influence of the compressive forcing on the intermittency in the Eulerian framework is an important result that needs further studies.\\
To estimate the influence of shocks and other non-local, inter-scale processes arising in supersonic, compressible turbulence, we compare our results with 
 the data of other subsonic, incompressible simulations.  
 Our scaling exponents are below those for incompressible media in both frameworks and are significantly different from the data of BBFLT10 and GFN02. 
 This indicates a more intermittent behaviour in our supersonic, compressible, turbulent flow, which is even stronger for the simulations with compressive forcing.
 The comparison of the scaling exponents of the tracer particles with the $\rho^{1/3}$ mass-weighted results of the grid in the Eulerian framework shows that
 the $\rho^{1/3}$ multiplier does not have the same effect as the averaging over tracer particles.
 The intrinsic mass-weighting of the tracer particles shows a less intermittent behaviour for both forcings than the results of SFK08.
\subsection{Intermittecy models for inertial range scaling}
\begin{figure}
\begin{center}
\includegraphics[width=1 \linewidth]{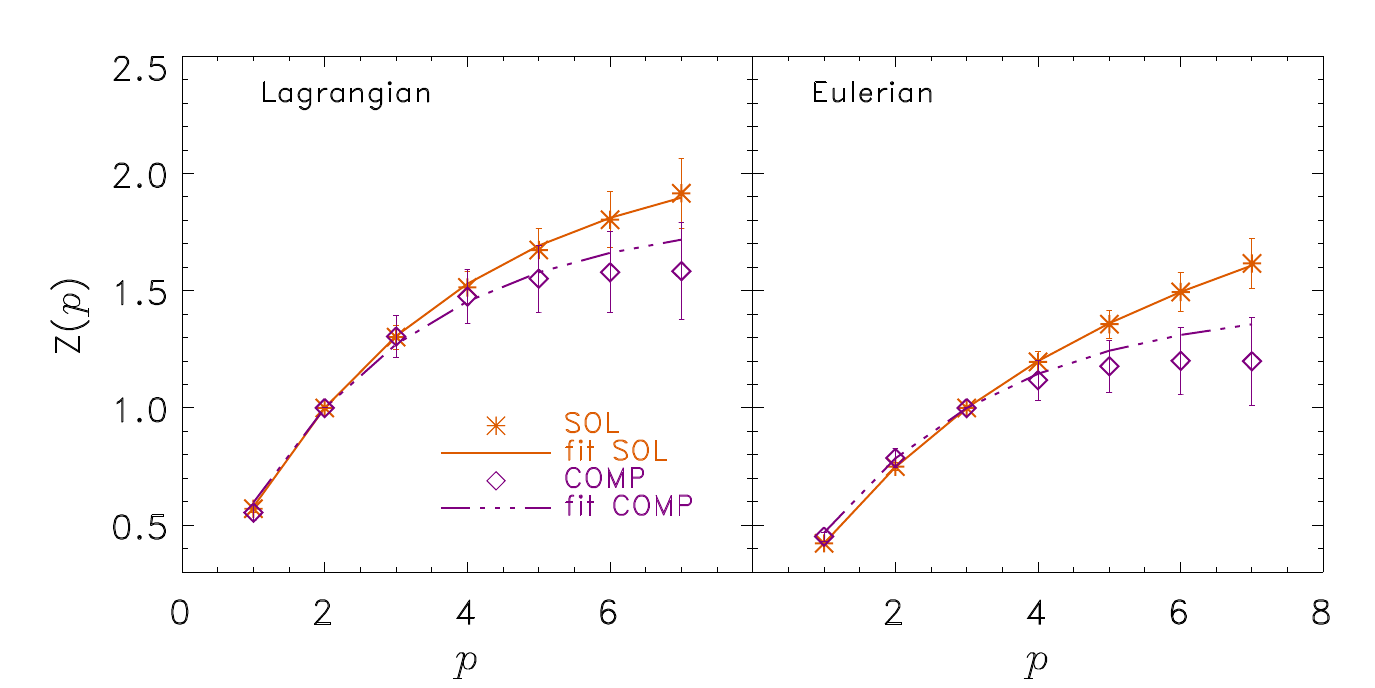}
\end{center}
\caption{Relative scaling exponents for the Lagrangian (left panel) and Eulerian (right panel) structure function for solenoidal and compressive forcing. The lines indicate
 a fit with the intermittency model proposed by \protect\citet{Dubrulle1994} with the assumption of \protect\citet{Schmidt2008}, $\Delta = 1$. In the Lagrangian framework, we get a
 dimension of the most dissipative structures $D_{L,\mathrm{sol}}=0.87$ and $D_{L,\mathrm{comp}}=1.17$ for solenoidal and compressive forcing respectively. In the Eulerian framework we get
 $D_{E,\mathrm{sol}}=1.11$ and $D_{E,\mathrm{comp}}=1.55$. The compressive forcing and the associated stronger density fluctuations cause a higher dimension of the most dissipative structures.}
\label{fig:Powerlaw}
\end{figure}

With the relative scaling exponents of the last section, we can compare our results with the predictions of intermittency models.
 We use the generalised equation of the phenomenological model of \citet{she94} introduced by \citet{Dubrulle1994} for the scaling exponents in the Eulerian framework
\begin{equation}
Z_E(p)=(1-\Delta_E)\frac{p}{3}+\frac{\Delta_E}{1-\beta_E}(1-\beta_E^{p/3}) \,.
\label{eq:she}
\end{equation}
With the assumptions $\tau \propto \ell^{2/3}$ and $LS^p \propto \langle \epsilon_{\tau}^{p/2} \rangle \tau^{p/2}$, where $ \langle \epsilon_{\tau}^{p} \rangle $ are the
 moments of the energy dissipation at the time scale $\tau$,
 one can show a similar equation for the Lagrangian
 framework, using the same arguments and derivations of \citet{she94}:
\begin{equation}
Z_L(p)=(1-\Delta_L)\frac{p}{2}+\frac{\Delta_L}{1-\beta_L}(1-\beta_L^{p/2}) \,.
\label{eq:sheL}
\end{equation}
For simplicity we use $\tau \propto \ell^{2/3}$ for the transformation into the Lagrangian framework, instead of $\tau \propto \ell$, and
 treat the influence of compressibility in both frameworks by having different values for $\Delta$ and $\beta$
 compared to the K41 theory \citep[see e.g.,][]{Boldyrev2002, Schmidt2008}.
 Figure~\ref{fig:Powerlaw} shows the measured scaling exponents and the fits with equation (\ref{eq:she}) and (\ref{eq:sheL}).
 For the fitting procedure we follow the idea of \citet{Schmidt2008, Schmidt2009} and
 set $\Delta=1$, which follows from Burgers scaling, $\tau \propto \ell$, as used in the last section, leaving us with only one free fitting parameter.
 With the measured $\beta$ we can calculate the co-dimension of the most dissipative structures $C=\Delta / (1-\beta)$, which is connected to the actual
 dimension of the most dissipative structures
 via $D=3-C$. The latter quantifies how volume-filling the most dissipative structures are in the turbulent medium.
 From our fits we get $D_{L, \mathrm{sol}} = 0.87$, $D_{L, \mathrm{comp}} = 1.17$ in the Lagrangian framework and
 $D_{E, \mathrm{sol}} = 1.11$, $D_{E, \mathrm{comp}} = 1.55$ in the Eulerian one. In the Eulerian framework, the most dissipative structures are between filamentary structures
 ($D=1$, as in \citet{she94}) and sheet-like structures
 ($D=2$, as proposed by \citet{Boldyrev2002} for the Kolmogorov-Burgers model). In the Lagrangian framework and for solenoidal forcing,
 the most dissipative structures are close to filamentary structures. Compressive forcing yields a larger fractal dimension
 than solenoidal forcing in both frameworks. Although the whole turbulent flow is more space filling for solenoidal forcing, as observed
 in figure~\ref{fig:DensVel}, the most dissipative structures of the compressive forcing have a larger dimension and are thus more space filling.
 However, it is unclear how to interpret these results in the
 one-dimensional Lagrangian framework of temporal increments rather than spatial increments as in the Eulerian framework.
 In the Eulerian framework, we can compare our results with the dimensions we get by calculating the scaling exponents with
 the mass-weighted velocities of the grid. \citet{Schmidt2008} measured $D_{\mathrm{sol}}=1.82$ and $D_{\mathrm{comp}}=1.92$, showing the same trend between the solenoidal and compressive
 forcing, but larger than the values we measured on the tracer particles. The reason for these differences is the more intermittent behaviour of the scaling exponents, as discussed above
 (see also table~\ref{tab:ScalingExponents}).
\section{A statistical theory of the large-scale velocity increments}
\label{sec:statisticaltheory}
\begin{figure}
\begin{center}
\includegraphics[width=1 \linewidth]{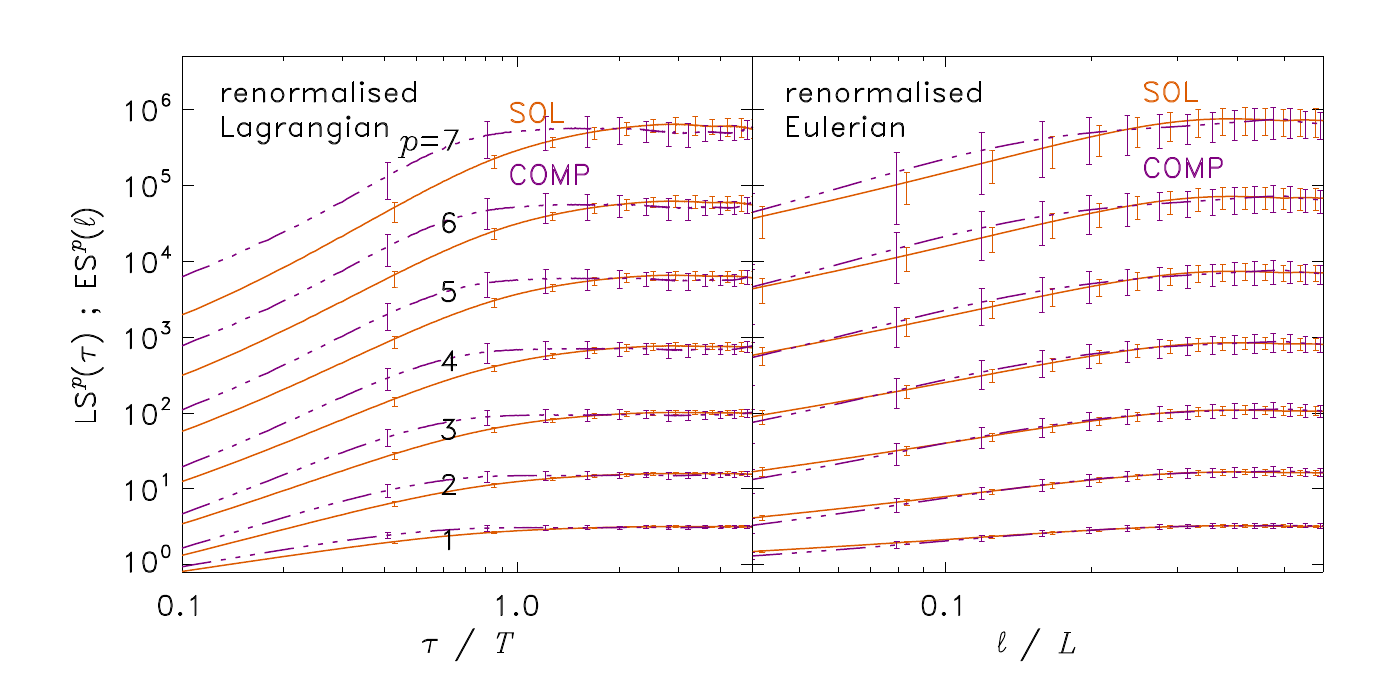}
\end{center}
\caption{Same as figure~\protect\ref{fig:Strukturfunktionen}, but with structure functions of the compressive
 forcing normalised to the r.m.s. Mach number of the solenoidal forcing, using our model prediction (\ref{eq:tollesDing_3}).
 The differences of the structure functions between the solenoidal and compressive forcing thus vanish in the integral range.}
\label{fig:norm_SF}
\end{figure}

\begin{figure}
\begin{center}
\includegraphics[width=0.7 \linewidth]{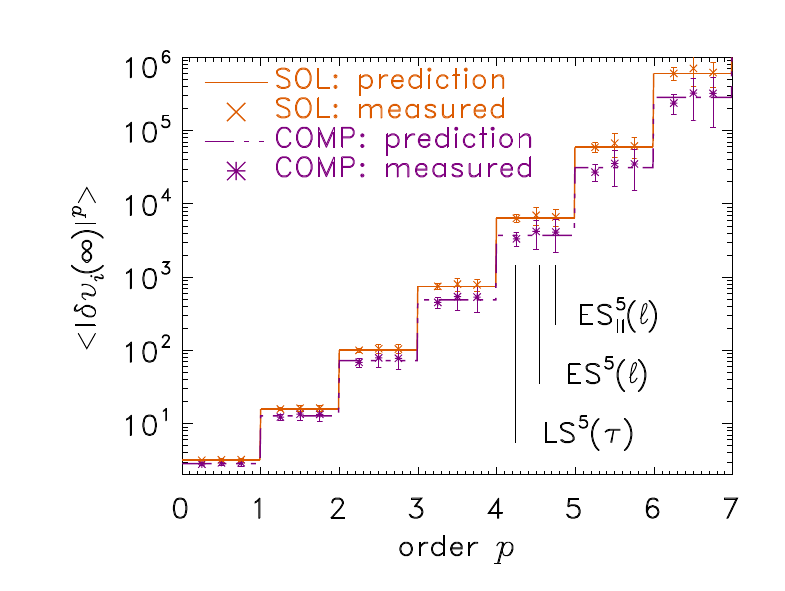}
\end{center}
\caption{Values of the saturated structure functions measured from the simulation (crosses, stars, and error bars) and the values predicted from formula~(\ref{eq:tollesDing_3})
 (solid and dashed-dotted lines)
 for different orders $p$. The values are shown to the left of their respective order and the crosses and stars are from left to right the values of the Lagrangian, Eulerian,
 and the longitudinal Eulerian structure function. The error bars of the measured saturation values are the averaged errors of the structure functions.}
\label{fig:Vorhersage_Beobachtung}
\end{figure}

In this section, we show that the statistical properties of the velocity increments in a turbulent flow
 on large scales can be described with only one parameter, the r.m.s Mach number. This is valid for
 velocity increments in the Lagrangian and Eulerian framework.
 The structure functions defined by (\ref{eq:LS}) to (\ref{eq:ES_para}) can be expressed as the moments of the PDFs of the velocity increment, which are functions
 of $\tau$ or $\ell$, so we can write for a general structure function,
 \begin{equation}
 S^p(\alpha) = \int{\left| \delta v \right|^p P( \delta v,\alpha) \mathrm{d}(\delta v)}\,,
 \label{eq:SF}
 \end{equation}
where $P( \delta v, \alpha)$ is the probability density of $\delta v$ with the increment $\alpha$.
 In the last section, we showed that the PDFs of the velocity increments converge towards a Gaussian distribution on the largest scales.
 The Gaussian form can be understood analytically as a consequence of the central limit theorem, assuming that the  two velocities,
 $v^m(\textbf{r}+\ell)$ and $v^n(\textbf{r})$ in space or $v^m(t+\tau)$ and $v^m(t)$ in time,
 are independent for large spatial or temporal increments.\\
With the Gaussian assumption, we can express the structure functions on large scales as
\begin{subeqnarray}
S^p(\alpha \rightarrow \infty) &=& \frac{2}{\sigma \sqrt{2 \pi }} \int\limits_{0}^{\infty} {( \delta v )}^p e^{- \frac{\left(\delta v\right)^2}{{2}\sigma^2} }\, \mathrm{d}\, (\delta v) \\
		               &=& \frac{\Gamma\left(\frac{p+1}{2}\right)}{\sqrt{\pi}}\,(\sqrt2\sigma)^p \, ,
\label{eq:tollesDing}
\end{subeqnarray}
where $S^p(\alpha)$ stands for any structure function of (\ref{eq:LS}) to (\ref{eq:ES_para}),
 $\alpha$ is the temporal or spatial increment, $\sigma$ is the standard deviation of the Gaussian distribution, and $\Gamma$ is the Gamma function.
 Equation~(\ref{eq:tollesDing}\textit{b}) describes the moments of the Rayleigh distribution, which is also the result for
 the moments of the total structure function with a velocity increment $\delta \mathbf{v} = \sqrt{\delta v_x^2 + \delta v_y^2 + \delta v_z^2}$,
 if the increments $\delta v_i$ follow a Gaussian distribution.\\ 
\citet{Stutzki1998} showed that 
\begin{equation}
 \langle {(\delta v(\ell\rightarrow \infty))}^2 \rangle =2 \mathcal{M}_M^2 c_{\mathrm{s}}^2\,,
 \label{eq:Stutzki1}
\end{equation}
where they used homogeneity and the fact that the autocorrelation vanishes for large spatial increments.
 In our case, the quantity $\mathcal{M}_M$ is a mass-weighted value, because the average in (\ref{eq:Stutzki1}) is taken over the velocity increments of the tracer particles.
 Furthermore, we assume that the second-order structure function is proportional to the kinetic energy for large increments
 and as the longitudinal structure function and the structure function averaged over the three directions of
 the coordinate system have only one-third degree of freedom compared with the total structure function,
\begin{equation}
\langle (\delta v)^2 \rangle = \langle \delta v_x^2 +\delta v_y^2 +\delta v_z^2 \rangle = 3 \langle (\delta v_i)^2 \rangle = 3 \langle (\delta v_{\parallel})^2 \rangle\,.
\end{equation}
If we combine this with (\ref{eq:tollesDing}) and (\ref{eq:Stutzki1}), we get a relation between the standard deviations of the Gaussian
 distributions and the r.m.s Mach number $\mathcal{M}_M$:
\begin{equation}
\langle (\delta v)^2 \rangle =2 \mathcal{M}_M^2 c_{\mathrm{s}}^2 =3 \sigma_i^2= 3 \sigma_{\parallel}^2 \,.
\end{equation}
The second-order moment can thus be used as a normalisation for our formula~(\ref{eq:tollesDing}) to predict the saturation level of the $p$th-order structure function
\begin{equation}
S^p(\alpha \rightarrow \infty) = \frac{\Gamma\left(\frac{p+1}{2}\right)}{\sqrt{\pi}}\,(\frac{2}{\sqrt{3}} \mathcal{M}_M)^p \, .
\label{eq:tollesDing_3}
\end{equation}
Figure~\ref{fig:norm_SF} shows the structure functions of figure~\ref{fig:Strukturfunktionen}, but renormalised with equation (\ref{eq:tollesDing_3})
 to the r.m.s. Mach number of the solenoidal forcing. The differences between the structure functions, driven by solenoidal and compressive forcing, vanishes in the integral range,
 what implies that the different forcings have no influence on the statistical properties of the structure functions in the integral range.
 Additionally, we verify this model by calculating the saturation behaviour with the measured r.m.s Mach number and compare the result with
 the saturation values extracted from figure~\ref{fig:Strukturfunktionen}. The result is summarised in figure~\ref{fig:Vorhersage_Beobachtung}.
 The measurements show an excellent agreement with the predicted values, for both solenoidal and compressive forcing.

\section{Summary and conclusions}
\label{sec:conclusion}
We have investigated the influence of solenoidal (divergence-free) and compressive (curl-free) forcing on the structure functions and density PDFs of a supersonic, compressible, turbulent flow using tracer particles in
 a set of three-dimensional numerical simulations.
 We analysed the density PDF, the PDFs of velocity increments, and the structure functions in the Lagrangian and Eulerian frameworks.
 As all of these quantities were  measured on tracer particles, we analysed mass-weighted statistics.
 Our main results and conclusions are:
\begin{itemize}
\item The solenoidal forcing yields a density PDF close to a log-normal distribution.
 In contrast, the compressive forcing yields distributions of the mass density that show stronger deviations from the log-normal shape in the wings of the distribution.
\item The compressive forcing excites stronger head-on collisions and shock fronts, which show a correlation between high density and low velocity, affecting the
 mass-weighted r.m.s Mach number, such that it becomes smaller than the volume-weighted Mach number. Similar holds for the solenoidal forcing, but the effect is weaker, as solenoidal forcing
 yields smaller density contrasts at the same r.m.s. Mach number.
\item The Lagrangian framework exhibits a more intermittent behaviour than the Eulerian one,
 measured with the deviations of the relative scaling exponents from the predicted intermittency-free K41 values
 and also with the kurtosis as an example for the higher moments of the PDF of the velocity increments.
 This analysis also shows that the turbulent medium, driven by the compressive forcing, is more intermittent than a medium, driven by solenoidal forcing.
 A comparison with simulations of incompressible turbulence shows that intermittency is stronger in a supersonic, compressible medium.
\item The influence of the different forcings are stronger in the Eulerian than in the Lagrangian framework, measured with the
 relative scaling exponents and the kurtosis of the velocity increments. 
\item The fractal dimension of the most dissipative structures are larger for compressive than for solenoidal forcing. The dimensions are around those of filamentary structures ($D=1$)
 in the Lagrangian framework and between filamentary and sheet-like structures ($1<D<2$) in the Eulerian framework.
 Although the whole turbulent flow driven by the solenoidal forcing is more space filling the most dissipative structures of the compressive forcing are more space filling.
\item The behaviour of the structure functions of all orders on the largest scales is determined by the r.m.s. Mach number of the system.
 With the assumption of a Gaussian distribution for the velocity increments on the largest scale, we derived an analytic relation, predicting
 the absolute values of the structure functions in the integral range of turbulence.
 The statistical properties of the velocity increments are indistinguishable in the integral range for both forcings types and follow our derived formula~(\ref{eq:tollesDing_3}).
 The different values of the saturated structure functions observed in
 figure~\ref{fig:Vorhersage_Beobachtung} are caused by the different Mach
 numbers
 ($\mathcal{M}_{\mathrm{sol},M} = 4.9 \pm 0.2 $ and
 $\mathcal{M}_{\mathrm{comp},M} = 4.4 \pm 0.3 $), and not by different
 statistical properties obtained by the different forcings. Thus, the predictions based on equation~(\ref{eq:tollesDing_3})
 are independent of the energy injection mechanism of the turbulence, but only depend on the mass-weighted r.m.s. Mach-number.
\end{itemize}
\begin{acknowledgments}      
 L.K.~acknowledges financial support by the International Max Planck Research School for Astronomy and Cosmic Physics (IMPRS-A) and the Heidelberg Graduate School of Fundamental
 Physics (HGSFP). The HGSFP is funded by the Excellence Initiative of the German Research Foundation DFG GSC 129/1.
 L.K., C.F., and 
R.S.K. acknowledge subsidies from the Baden-W\"{u}rttemberg-Stiftung via the program {\em Internationale Spitzenforschung II} (grant P-LS-SPII/18)
 from the German Bundesministerium f\"ur Bildung und Forschung via the ASTRONET project STAR FORMAT (grant 05A09VHA).
 C.F.~has received funding from the European Research Council under the European Community's Seventh Framework Programme (FP7/2007-2013 Grant Agreement no.~247060)
 and from a Discovery Projects fellowship by the Australian Research Council (no. DP110102191) 
 for the research presented in this work.
 R.S.K. furthermore gives thanks for subsidies from the Deutsche Forschungsgemeinschaft (DFG) under grants KL 1358/10, and KL 1358/11
 and via the SFB 881 'The Milky Way System',
 as well as from a Frontier grant of Heidelberg University sponsored by the German Excellence Initiative.
 Supercomputing time at the Leibniz Rechenzentrum
 (grant no.~h0972 and pr32lo), and at the Forschungszentrum J\"ulich (grant no.~hhd20) are gratefully acknowledged. The software used in this work was in
 part developed by the DOE-supported ASC / Alliance Center for Astrophysical Thermonuclear Flashes at the University of Chicago.
\end{acknowledgments}
\appendix
\section{Influence of the numerical grid resolution}
\label{sec:resolutionstudy}
\begin{figure}
\includegraphics[width=1 \linewidth]{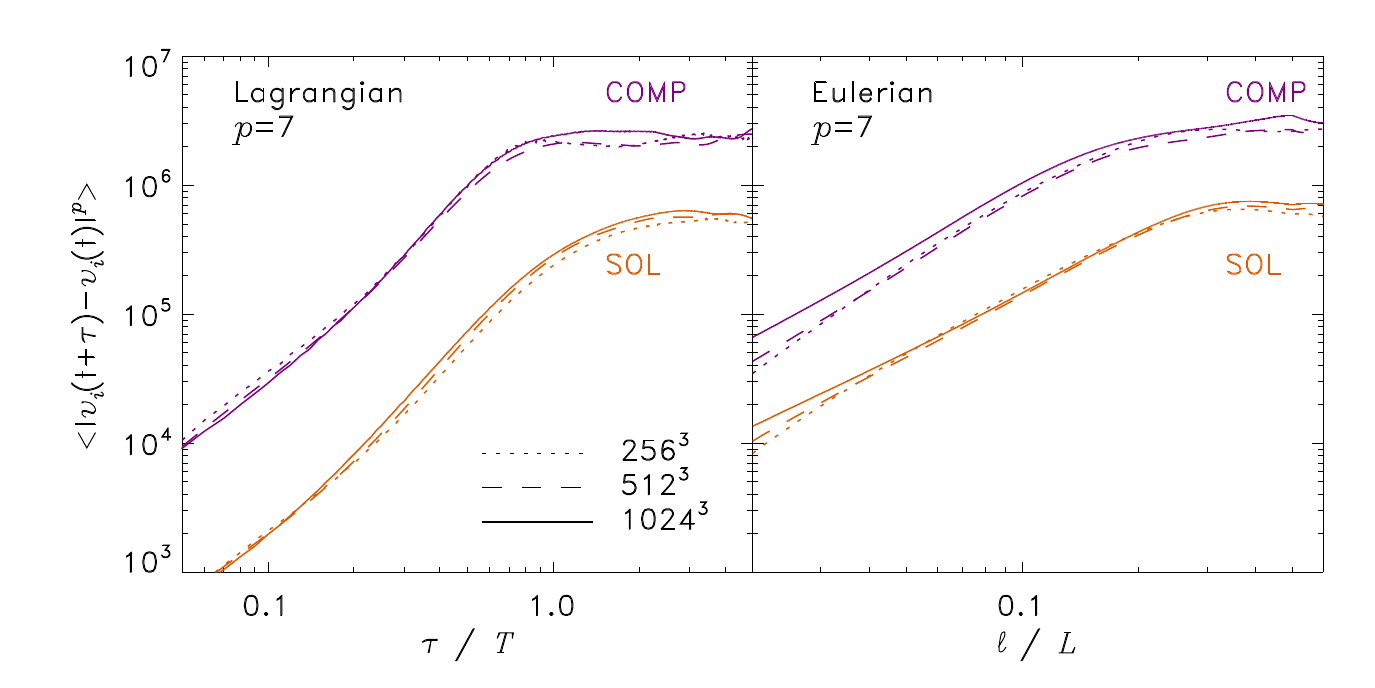}
\caption{Structure functions of order $p=7$ for different grid resolutions, shown for both forcing types.
 The LSF (left panel) with a low grid resolution of $256^3$ and $512^3$ were calculated with $128^3$ and $256^3$ tracer particles and the simulation with $1024^3$ grid cells was calculated with $5\times10^6$ tracer particles.
 The ESF (right panel) was calculated with $16^3$ sub boxes and $10^{10}$ sampling pairs. The structure functions of the compressive forcing was multiplied
 with a factor of $10$ so that the structure functions are distinguishable.}
\label{fig:Resolution}
\end{figure}

 Figure~\ref{fig:Resolution} (left panel) shows the LSF of order $p=7$ for both
 forcing types and with the grid resolutions $256^3$,
 $512^3$, and $1024^3$. The right panel shows the same for the ESF. The LSFs are calculated with $128^3\approx 2.1$,
 $512^3\approx 16.8$ and $5$ million tracer particles
 for the different grid resolutions, respectively.
 The ESFs are calculated with $16^3$ sub boxes and with $10^{10}$ sampling pairs.
 Figure~\ref{fig:Resolution} shows that the structure functions of order $p=7$ differ by about $15\%$,
 caused by the different grid resolutions.
 This is of the same order as the $1\,\sigma$ variations in time of the structure functions indicated as error bars
 in figure~\ref{fig:Strukturfunktionen}. Therefore, the influence of the resolution is smaller than the temporal variations.
\section{Convergence test for the structure functions}
\label{sec:ConvergenceESF}
In order to verify that our calculations are converged with a sufficient number of data pairs to sample the structure functions, we show that 
 the structure functions do not change significantly by further increasing the number of sampling pairs.
 As large velocity fluctuations have a stronger influence on the higher orders of the structure functions and these events are very rare, the
 statistical convergence of the higher orders is slower compared with the lower orders. Thus, if we can demonstrate convergence for the higher order structure functions,
 this automatically holds for all lower orders.
\begin{figure}
\begin{center}
\includegraphics[width=1 \linewidth]{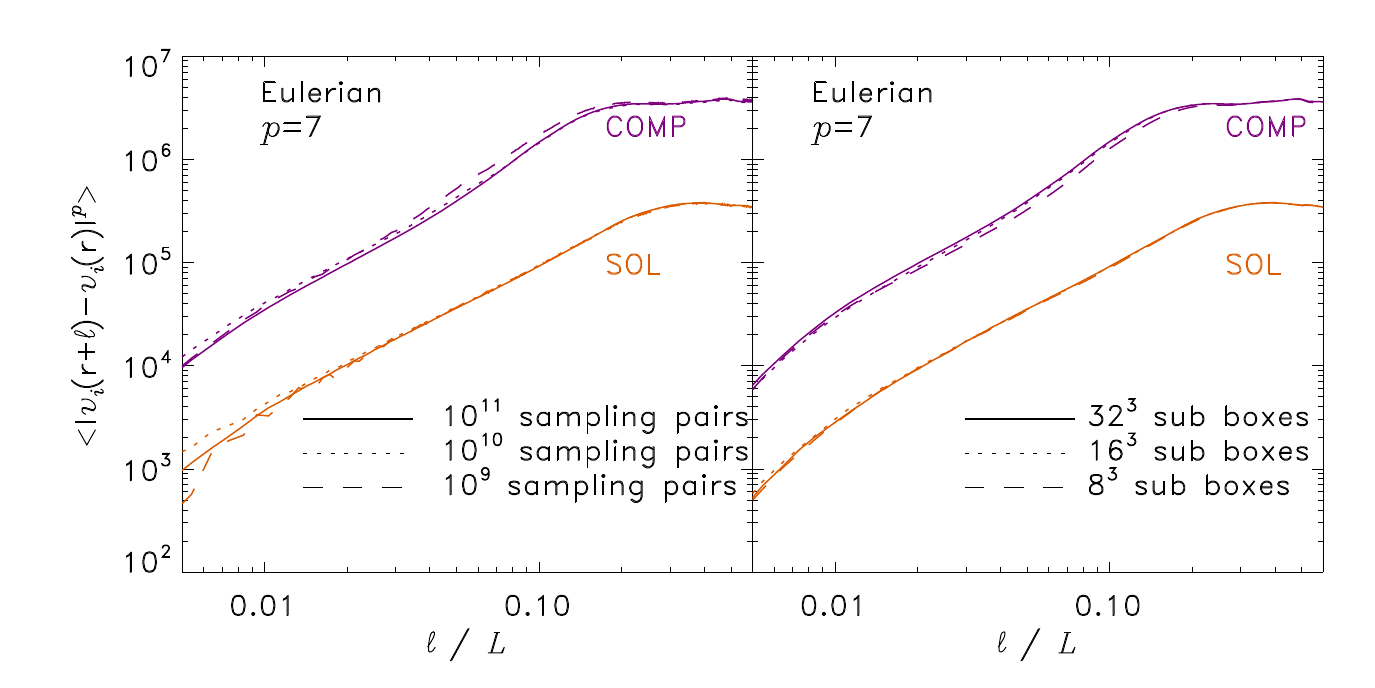}
\end{center}
\caption{Left panel: ESF of order $p=7$ for both forcing types, $16^3$ sub boxes and different numbers of sampling pairs.
 The structure functions are converged on large scales for more than $10^{10}$ sampling pairs.
 Right panel: Same as left panel, however with $10^{11}$ sampling pairs and different numbers of sub boxes. In both figures, the structure functions of the compressive forcing was multiplied by
 a factor of $10$.}
\label{fig:Konvergenz_Euler}
\end{figure} 
Figure~\ref{fig:Konvergenz_Euler} (left panel) shows the Eulerian structure function of order $p=7$ for solenoidal and compressive forcing. The structure function of the compressive
 forcing is multiplied by a factor of $10$, so that the structure functions of the different forcings are distinguishable.
 In order to check the convergence, we use one random time sample ($t=4T$) in the state of fully-developed turbulence, $16^3$ sub boxes
 and different numbers of sampling pairs ($10^9$, $10^{10}$, $10^{11}$). Increasing the number of sampling pairs further only influences
 small scales, $\ell < 0.07\,L$.
 The structure functions are converged on larger scales.
 For the Eulerian structure function, we also verified that the method of selecting tracer particles for the calculation with our procedure of sub boxes has no
 significant influence on the results. Therefore, we calculated the structure function with $10^{11}$ sampling pairs and different numbers of sub boxes ($8^3$, $16^3$ and $32^3$).
 Figure~\ref{fig:Konvergenz_Euler}  (right panel) shows that further increasing the number of sub boxes also only influences scales
 $\ell < 0.05\,L$, and with $16^3$ sub boxes, the structure functions are converged.\\
For the Lagrangian structure function, we also have to verify that the structure functions do not change significantly with the number of sampling pairs. We 
 calculate the LSF for all $512^3$ tracer particles for one time-line from $t=2\,T$ to $t=10\,T$ and compare it with the LSF calculated with $5$ and $10$ million tracer particles.
 The results are shown in figure~\ref{fig:Lag_convergence}, where the structure functions with compressive forcing are multiplied with a factor of $10$.
\begin{figure}
\begin{center}
\includegraphics[width=0.7 \linewidth]{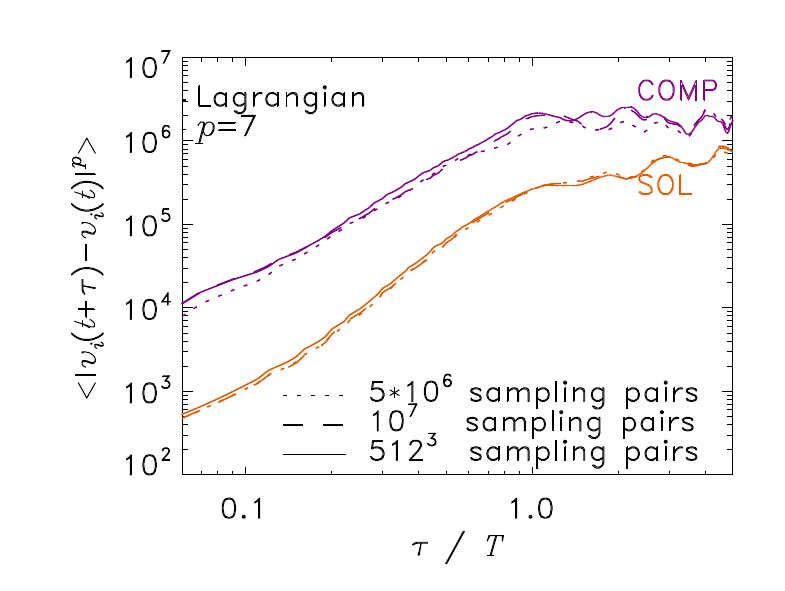}
\end{center}
\caption{LSF of order $p=7$ for both forcing types and $5$, $10$, and $512^3 \approx 134$ million sampling pairs.
 The structure functions calculated with solenoidal forcing are converged on
 all scales. The structure functions calculated with compressive forcing show a small influence of the number of used sampling pairs.
 The structure function of the compressive forcing was multiplied by a factor of $10$.}
\label{fig:Lag_convergence}
\end{figure} 
 The structure functions calculated for solenoidal forcing are converged on
 all scales, and the structure functions calculated for compressive forcing show only small variations with the number of sampling pairs.
 The reason for the large fluctuations in the integral range in figure~\ref{fig:Lag_convergence} is that
 the LSF was here calculated with one time-line only. 
 Figure~\ref{fig:Lag_convergence} shows that the time evolution of the forcing module has a direct influence on the amplitudes of the velocity increments
 in the integral range, but these fluctuations are smaller than the variations in time, we use as errors in figure~\ref{fig:Strukturfunktionen}.
 However, this direct influence vanishes on average by using different staring times for calculating the LSF.
 In the inertial range with $\tau < 1\,T$, the structure functions in figure~\ref{fig:Strukturfunktionen} have a factor of about $700$ more sampling pairs for each bin.
 This large statistic we used there ensures that our structure functions are also converged in the inertial range.
\bibliographystyle{jfm}
\bibliography{mybib}
\end{document}